\documentclass[twocolumn,showpacs,preprintnumbers,amsmath,amssymb]{revtex4}

\usepackage{graphicx}
\usepackage{dcolumn}
\usepackage{bm}

\begin{document}

\title{Decoupled two-dimensional superconductivity and continuous melting transitions \\
in layered superconductors immersed in a parallel magnetic field}

\author{Xiao Hu}
\author{Masashi Tachiki}
\affiliation{Computational Materials Science Center,
National Institute for Materials Science, Tsukuba 305-0047, Japan}

\date{May 14, 2004}
%\date{March 18, 2004}
%\date{August 22, 2003}

\begin{abstract}
Possible phases and the \(B-T\) phase diagram of interlayer Josephson vortices induced by a magnetic
field parallel to the superconducting layers are investigated 
by Monte Carlo simulations based on the anisotropic, frustrated XY model.   While for
low magnetic fields and small anisotropy parameters a single first-order transition is observed similarly 
to the melting of Abrikosov (or pancake) vortex lattice, an intermediate phase, characterized by decoupled, 
two-dimensional (2D) quasi long-range crystalline order (QLRCO) and superconductivity, is found at high magnetic
fields and large anisotropy parameters.   Combining the simulation results with a symmetry argument, 
it is revealed that this intermediate phase is of Kosterlitz-Thouless (KT) type, and the melting of 
2D quasi Josephson vortex lattices and suppression of superconductivity is a KT transition.  Evolution 
of the intermediate phase to the low-temperature phase of 3D LRCO  is second order and 
belongs to the 3D XY universality class.  The three phase boundaries merge at a multicritical 
point at the magnetic field of order \(B_{\rm mc}=\phi_0/2\sqrt{3}\gamma d^2\) in the \(B-T\) phase diagram.
It is revealed that decoupling of the 3D Josephson vortex lattice into the 2D phase is triggered by hops of 
Josephson flux lines across superconducting layers activated by thermal fluctuations.  
The equilibrium phase diagram with 
the KT phase at high magnetic fields and large anisotropy parameters is consistent with the peculiar 
Lorentz-force-independent dissipation observed in highly anisotropic high-\(T_c\) superconductor 
Bi\(_2\)Sr\(_2\)CaCu\(_2\)O\(_{8+y}\) by Iye {\it et al.} (Physica {\bf 159C}, 433 (1989)).

\end{abstract}


\maketitle

\section{Introduction}

The high-\(T_c\) superconductors share a layered structure in which the superconductivity 
is widely believed to occur mainly in the CuO\(_2\) layers intervened by layers of charge 
reservoir (block layers).   For many purposes, the materials can be considered as intrinsic
stacks of superconducting layers, coupling with each other by the Josephson effect.  They are
therefore strongly anisotropic between the normal (\(c\)-axis) and in-plane directions.
In spite of this anisotropy, the superconductors are three-dimensional (3D) in nature 
at temperatures near the superconductivity transition point at zero magnetic field \( T_c\). 
Especially, the critical phenomena of superconductivity transition are 
governed by the 3D XY universality class \cite{Kamal}.  Since the coherence lengths diverge in
the critical region, the short-range (SR) nonuniformity caused by the layer structure becomes
unimportant.  

There are two cases that physics can be different.  The first case is realized at low temperatures,
at which the correlation length \(\xi_c\) becomes comparable with, and even
smaller than the interlayer separation \(d\), where two-dimensional (2D) properties can arise.  A 
crossover between 2D and 3D behaviors is expected to occur when the value of \(\xi_c/d\) is of the 
order of unity.  The second case is realized when the superconductor is immersed in a strong
magnetic field parallel to the CuO\(_2\) layers.  The coupling between the
superconducting layers are weakened significantly by the magnetic field, and the effective 
correlation length \(\xi_c\) 
can be very small even at high temperatures.  To elucidate the second case is the objective of 
the present study.  As revealed in the present paper, there is a genuine thermodynamic phase transition 
between 2D and 3D phases at high magnetic fields.

It is now well established that the superconductivity transition in a type II 
superconductor in magnetic fields \cite{Abrikosov}  is first order, accompanied by
the freezing of the flux-line liquid into 3D lattice \cite{Blatter1,Crabtree,Nattermann}.  This notion is 
particularly important because the transition was considered second order for a long time.   We notice however
that the first-order normal to superconductivity transition is mainly observed under magnetic fields perpendicular
to the layers, where the 2D translation symmetry enjoyed by pancake
vortices is broken at the transition. In a sharp contrast, a parallel magnetic field penetrates the layered 
superconductor
in the form of Josephson vortex of flux quantum \cite{Dolan} through block layers \cite{Tachiki}.  The relevant
\(c\)-axis translation symmetry is reduced {\it a priori} to a discrete one, raising the possibility of
new phases and new melting process. 

As an order competing with the 3D lattice order, decoupled superconducting layers with in-plane 
quasi-long-range (QLR) correlations was proposed by Efetov even before the discovery of 
high-\(T_c\) superconductors with profound layer structures \cite{Efetov}.  
Later on, an exponential dependence of the interlayer shear modulus on the magnetic field is
derived by Ivlev {\it et al.} \cite{Ivlev}, consistently with the decoupling scenario since the small shear modulus 
can hardly
hold the 3D lattice when thermal fluctuations set in.  This possibility was however questioned by Mikheev and 
Kolomeisky 
\cite{Mikheev} using the renormalization group (RG) approach.  It is derived that any weak interlayer
coupling should be relevant and only a 3D LR crystalline order (CO) is possible, as far as 
hops of Josephson flux lines across CuO\(_2\) layers can be neglected, an approximation adopted in the previous works.
A similar conclusion was obtained by Korshunov and Larkin \cite{Korshunov}.   

Transport experiments on high-\(T_c\) superconductors have been sheding lights on the possible crystalline
order of Josephson flux lines.  A peculiar {\it Lorentz-force-independent} dissipation has 
been found in Bi\(_2\)Sr\(_2\)CaCu\(_2\)O\(_{8+y}\) under \(H=5 T\) by Iye {\it et al.}
that the resistivity is independent of the angle between the 
magnetic field and current when they are both parallel to the CuO\(_2\) layer \cite{Iye}. 
In the same family of materials, non-Ohmic power-law I-V characteristics are observed by 
other groups \cite{Ando,Pradhan}.   
Chakravarty {\it et al.} demonstrated that the {\it Lorentz-force-independent} dissipation
cannot occur in a lattice phase \cite{Chakravarty}.  In order to explain these
peculiar dissipations, Blatter {\it et al.}  proposed a Kosterlitz-Thouless
 (KT) transition \cite{Berezinsky,KT}  at high magnetic fields, relating the creeping of 
Josephson flux lines across the CuO\(_2\) layers caused by applied currents with the shearless 
flux-line state \cite{Blatter}.  (See \cite{Horovitz} for a possible KT phase at 
intermediate magnetic fields.)   On the other hand,
an excess resistivity for \(H\perp I\) over \(H|| I\), and thus a {\it Lorentz-force-dependent} 
dissipation was observed in YBa\(_2\)Cu\(_3\)O\(_{7-\delta}\) single crystals at low fields
\cite{Palstra,Kwok1,Budhani}.   Kes {\it et al.} addressed this difference in 
terms of the different anisotropy parameters \cite{Kes}.  Taking the Bi-based materials almost
"magnetically transparent", the authors attributed the experimentally observed
dissipations to uncontrolled miss alignments of magnetic field. 

Motivated by an experiment by Kwok {\it et al.} suggestive of continuous melting 
transition at intermediate magnetic fields \cite{Kwok}, Balents and Nelson proposed a
smectic phase: a regular subset of block layers are selectively occupied by Josephson 
vortices; in the occupied block layers Josephson flux lines behave like liquid.
They showed that melting of the 3D lattice can be continuous in the presence of layer pinning \cite{Balents}.  
See also \cite{Radzihovsky}. 

Although these studies provide important physical insights for the mixed states in layered type II 
superconductors, it is clear that discrepancies among different approaches have not been 
resolved and that a unified picture is still not available. The main difficulty lies in the features
of different energy contributions involved in the present system:  First, 
the inter-vortex repulsive force is highly anisotropic between the \(c\) axis and the \(ab\) plane; 
Second, Josephson flux lines feel strong pinning potentials from the CuO\(_2\) layers.
 At a first glance, approximations which take into account differences in strengths of these energy 
contributions can simplify the problem.  It turns out that the situation is quite complicated.  
The Josephson flux lines arrange themselves 
such that they reside at positions from which deviations in different directions cost 
equal energies.   Although the profound layer structure reduces the \(c\)-axis component of the 
displacement field at low temperatures, it certainly enhances the rigidity of the flux-line lattice 
and stabilizes it to high temperatures.  At the melting temperature, thermal fluctuations are
significant and smear out the strong layer pinning to an effectively weak one.  Therefore,
 no approximation can be justified easily in the regime of phase transition.  One thus has to
treat the competition among the anisotropic repulsion energy, the periodic layer pinning potential, 
and thermal fluctuations simultaneously.  

On the other hand, since a Josephson vortex is of large size in the in-plane direction, point-like
defects play much less pinning effect compared with a pancake vortex.  It relieves one from being
involved into glassy problems, and makes the study on a pure Josephson-vortex system more
relevant to reality.  As a result, the high-field part of the \(B-T\) phase diagram for a pure vortex
system becomes accessible experimentally.  For this point, please see several interesting proposals
\cite{DHLee,Frey,Glazman}  in terms of pancake vortices, which are however hampered by inevitable 
point-like defects in superconductors which govern the behaviors of pancake vortices.

These situations motivate us to investigate Josephson-vortex systems
by means of Monte Carlo simulations.  The Hamiltonian is the so-called anisotropic,
frustrated XY model on the superconductivity order parameter, where the frustrations in
phase variables are induced by the magnetic field.  This model has been used to simulate quite successfully
the melting phenomenon of Abrikosov (or pancake) vortex lattice under magnetic
fields parallel to the \(c\) axis \cite{Hetzel,Teitel,Huc,Koshelev,Nguyen,Olsson}.   We adopt
the same model in the present study on interlayer Josephson vortices.  The main results are
summarized as follows.  There is a multicritical point in the \(B-T\) phase diagram: 
below the multicritical field a single first-order melting transition upon temperature sweeping is
observed; above it there exists an intermediate KT phase 
characterized by in-plane 2D, QLRCO and superconductivity in between the normal phase and 3D lattice phase,
accompanied by two continuous melting transitions.  The existence of a KT phase at magnetic 
fields above the multicritical point explains the peculiar dissipations,
such as the {\it Lorentz-force-independent} resistance and the power-law non-Ohmic I-V characteristics,
observed in Bi-based materials for which
the multicritical magnetic field is approximately \(B_{\rm mc}\simeq 2T\) presuming \(\gamma=150\).
In contrast, Lorentz-force-dependent dissipations are expected for 
YBa\(_2\)Cu\(_3\)O\(_{7-\delta}\), since to access the KT phase one needs  
a magnetic field above \(B_{\rm mc}\simeq 50T\) taking \(\gamma=8\).  Some of the 
results were published in Ref.\cite{Huab0,Huab}.

The remaining part of this paper is organized as follows.  In Sec. II the model Hamiltonian is 
introduced and some details of the simulation techniques are presented.  After a discussion on 
the description on flux line lattice, we show in Sec. III simulation results on the first-order melting 
for a system corresponding to YBa\(_2\)Cu\(_3\)O\(_{7-\delta}\) single crystals.
It then follows numerical evidences on an intermediate phase in a highly anisotropic system.
Detailed characteristics are provided which allow us to conclude the KT nature of this phase.  The
superconductivity transition is discussed in Sec. IV, and the relationship with the crystallin order of Josephson
vortices is revealed. Section V addresses the possible universality of the transition between 2D phase and 
the 3D lattice. The \(B-T\) phase diagram of Josephson vortex systems is mapped out in Sec. VI.  Finally, 
summary on the main results derived from simulations and discussions on recent experiments are given in Sec. VII.

\section{Model and simulation details}

In the presence of magnetic field, the amplitude of superconductivity order parameter, associated 
with the local tendency of electron pairing, attains a finite value at \(H_{c2}\) through a crossover.  The genuine 
thermodynamic phase transition takes place at a temperature lower than that corresponding to \(H_{c2}(T)\). 
Therefore, the most important thermal fluctuations in the thermodynamic phase transition come from 
phase variables of the superconductivity order parameter.  Under a magnetic field parallel to the layers,
the effective Hamiltonian is thus \cite{Huab}

\begin{equation}
\hspace{-30mm} {\cal H}=  -J\sum_{ {\bf R}_i-{\bf R}_j \parallel x,y \hspace{0.5mm} {\rm axis}} 
                     \cos(\varphi_i -\varphi_j)  
\nonumber
\end{equation}

\vspace{-5mm}

\begin{equation}
  -\frac{J}{\gamma^2}\sum_{ {\bf R}_i-{\bf R}_j  \parallel c \hspace{0.5mm} {\rm axis}} 
    \cos(\varphi_i-\varphi_j
                 -\frac{2\pi}{\phi_0} \int^{{\bf R}_j}_{{\bf R}_i}{\bf A}\cdot d{\bf R}).
\end{equation}

\noindent The model is defined on the simple cubic grid with the unit length equal to the separation between 
CuO\(_2\) layers \(d\).  The couplings given by \(J=\phi^2_0 d/16\pi^3\lambda^2_{ab}\) are limited to nearest 
neighboring grid sites and  \(\gamma=\lambda_c/\lambda_{ab}\). 
To be specific, we put \(\hat{x}\perp \hat{y}\perp \hat{c} \) and  \({\bf B} || \hat{y}\), and choose the
Landau gauge \({\bf A}=(0, 0, -xB)\) in Eq.(1).   

The cosine functions of the gauge invariant phase differences in \(ab\) planes model the
kinetic energy terms in a Ginzburg-Landau free energy functional.  Higher harmonics are
neglected since the lowest order terms with \(2\pi\) modulation are sufficient for
describing relevant thermal fluctuations at phase transitions.  The second term in the Hamiltonian
is the Josephson energy between neighboring CuO\(_2\) layers.  The magnetic 
field weakens the interlayer coupling by inducing frustrations. Our Hamiltonian can be derived from
the Lawrence-Doniach free energy functional \cite{LD}, which is 
proposed for layered superconductors and recognized as a good model for high-\(T_c\)
superconductors with profound layer structure. 

Spatial variations and thermal fluctuations in the magnetic induction, and thus magnetic couplings, 
are neglected in Hamiltonian (1),
which is in principle justified only at the strongly type II limit.  It is a reasonable approximation
in the present case.  The London penetration depth \(\lambda_c\) is very large (of order \(\mu m\)),
which makes the spatial variation of the magnetic induction in the \(x\) direction and thus the
magnetic interactions among Josephson flux lines very small.   Magnetic couplings may play a role
when  pancake vortices are activated by thermal fluctuations.  It was addressed,  however,
that the magnetic interaction between two pancake vortices on a same CuO\(_2\) layer increases with
the separation only as \(1/R\) at large distance because of the interlayer coupling
 \cite{Blatter}, which is therefore much smaller than the elastic energy.   Therefore, magnetic couplings 
can be ignored safely as far as  the thermodynamic phase transition is concerned. 

Thermal fluctuations in the amplitude of superconductivity order parameter can be taken into
account in several ways.  One can include a temperature dependence of the amplitude in the mean-field 
way, such that the penetration depth and the coupling constants vary with temperature.
This treatment permits a quantitative comparison between simulated quantities and experimental
observations \cite{Dodgson,Huc}.
Alternatively, one can take thermal averages of superconductivity order parameter on scales larger
than the unit length of grid in \(ab\) planes.   A fully disordered configuration of phase variables on the
scale of \(d\) then results in zero superconductivity order parameter at any larger length scales.

The Hamiltonian (1) provides a reasonable description for the layer pinning on Josephson vortices in
the whole temperature regime, even without taking into account the temperature dependence of 
Ginzburg-Landau correlation length. At low temperatures, the layers modeled by the discreteness of the underlying
grid in the \(c\) direction set a series of barriers to the motion of Josephon flux lines.  If temperature 
is high enough, on the other hand, pancake vortices are activated thermally on CuO\(_2\) layers, 
such that Josephson flux lines can hop to neighboring block layers easily.  Thermal fluctuations
reduce effectively the layer pinning potential. 

The above Hamiltonian has a trivial limit \(\gamma=\infty\), where KT transition
takes place at \(T_{\rm KT}^{\rm bare}\simeq 0.89J/k_B\) in each independent \(ab\) plane.
For large but finite anisotropy parameters, there is a broad crossover regime which is dominated by this
limit, which can be seen in the specific heat \cite{Huab,Holme}.  

In the present study, the magnetic field is fixed at
\(f\equiv Bd^2/\phi_0=1/32\) while the anisotropy parameter is tuned for the convenience of 
simulation.  The system size is \(L_x\times L_y\times L_c=384d\times 200d\times 20d\) under 
periodic boundary conditions.  There are totally \(N_v=48000\) Josephson vortices in the ground    
state.  The system size is set anisotropically since we are interested in
systems of large anisotropy parameters:  there are \(f\times L_x/d=12\) Josephson flux lines the \(x\) direction,
which is comparable with the number of Josephson flux lines in the \(c\) axis: \(L_c/2d=10\).
(See Fig. 2 of \cite{Huab}.)   As the field direction coincides with \(ab\)
planes of strong couplings, the system size in this direction has to be taken large enough 
in order to treat thermal fluctuations sufficiently.  Although a full analysis on finite-size effects
is still difficult with the available computing resource, it will be revealed later 
 by comparing simulation results for different anisotropy parameters that the phases and phase diagram
derived in the present paper are free of serious finite-size effects.

In a typical simulation process, we start with a random configuration of phase variables
at a sufficient high temperature.  The system is then cooled down gradually 
with the temperature skips and the number of Monte Carlo sweeps
listed in Ref.\cite{Huab}.  After arriving at the lattice phase at a sufficiently low temperature, 
\(T=0.1J/k_B\), we heat the system back slowly.  Thermal averages are taken over \(\sim 10^7\)
Monte Carlo sweeps at temperatures around the transition points. All the results shown in the present
paper are calculated in the heating process.   

\section{Crystallization transition}

\subsection{Description of Josephson vortex lattice}

The crystalline order of the system is described by the correlation function

\begin{equation}
S( {\bf R} ) = \langle \rho ( {\bf R} ) \rho ( 0 ) \rangle 
                 - \langle \rho ( {\bf R} )  \rangle \langle \rho ( 0 ) \rangle, 
\end{equation}

\noindent where \({\bf R}=(x,y,z)\) and \(\rho({\bf R})\) is the \(y\) component
of the vorticity at position \({\bf R}\) which is explicitely defined by
\(\rho({\bf R} )\equiv \sum_{\alpha}\delta({\bf R}_{\perp}-{\bf R}_{\perp,\alpha}(y))\) 
with \({\bf R}_{\perp}=(x,z)\),
and its Fourier transformation, {\it i.e.} the structure factor,

\begin{equation}
S( {\bf k} )  = \int d^3R e^{ -i {\bf k}  \cdot {\bf R} }  S( {\bf R}  ).  
\end{equation}

In lattice phase subject to thermal fluctuations, the vortex density is expressed
in terms of  the reciprocal vectors as

\begin{equation}
\rho ( {\bf R} ) = \Psi_0 e ^{ -i {\bf K}_0 \cdot {\bf R} } 
                     + \Psi_1(  {\bf R} ) e ^{ -i {\bf K}_1 \cdot {\bf R} }  
                     + \Psi_2(  {\bf R} ) e ^{ -i {\bf K}_2 \cdot {\bf R} }  + c.c.,
\label{eqn:density}
\end{equation}

\noindent where \({\bf K}_j\) with \(j=0,1,2\) are primitive reciprocal lattice vectors and
\( {\bf K}_0+ {\bf K}_1+ {\bf K}_2=0\).  Higher harmonics are
not included for simplicity, which are not important for the later discussions on
phase transitions.   \( \Psi_{j}(  {\bf R} ) \)'s are order parameters of the crystalline
order, and are expressed by the displacement field \({\bf u}=(u_x,0,u_c)\) \cite{Landau}:

\begin{equation}
\Psi_j(  {\bf R} ) =  e ^{ i {\bf K}_{j} \cdot {\bf u} (  {\bf R} ) }.  
\label{eqn:op}
\end{equation} 

The density-density correlation is then given by correlation functions
of order parameters 

\begin{equation}
\begin{array}{rl}
S( {\bf R} ) & = \sum_{j}[e^{ -i {\bf K}_j  \cdot {\bf R} }  
                      \langle  \Psi_j(  {\bf R} )  \Psi_j^*( 0 )  \rangle + c.c. ]  \\[10pt]
   & = \sum_{j} |\langle \Psi_j \rangle |^2  [ e^{ -i {\bf K}_j  \cdot {\bf R} }  + c.c. ]  g( {\bf K}_j, {\bf R}), 
\end{array}  
\end{equation}

\noindent with the correlator of displacement fields defined by

\begin{equation}
g( {\bf K}, {\bf R}) \equiv \langle  e^{-i{\bf K}\cdot [ {\bf u}({\bf R}) - {\bf u}(0) ] } \rangle.
\end{equation}

\noindent  The cross terms between different reciprocal vectors disappear because of the
infinite expectation values of displacement fields.  The structure factor is then given by

\begin{equation}
S( {\bf k} ) \sim \sum_{j}   
                  \int d^3R  [ e^{ -i ( {\bf k} - {\bf K}_j )   \cdot {\bf R}  }  
                                 +e^{ -i ( {\bf k} + {\bf K}_j )   \cdot {\bf R}  }  ]  g( {\bf K}_j, {\bf R}).
\end{equation}

For the present flux line system, it is convenient to consider partial structure factors in the section
perpendicular to the magnetic field.  They are related to the above 3D one by a partial
integral, \( S(k_x, y, k_c)\sim \int dk_y e^{ik_y y}S({\bf k})\).  In the present paper, we will show
simulation results on \( S(k_x, k_c)\equiv S(k_x, y=0, k_c)\) which reveals the vortex correlations
in the same crosssection perpendicular to the magnetic field.

In a 3D crystal phase, \(g( {\bf K}_j, {\bf R})\)'s approach to constants at large distances, 
and \(\delta\)-function Bragg peaks should be observed at the reciprocal-lattice vectors.  
For the present system of weak interlayer couplings, there is a competing order to the 3D crystalline
order, in which the Josephson vortices are strongly correlated in each block layer such
that a QLRO is present, while only SR correlations are realized in the \(c\) direction:

\begin{equation}
g( {\bf K}_j, {\bf R}) \sim e^{ -|z|/\xi_c}/r^{\eta}
\label{eqn:rscf}
\end{equation}

\noindent with \({\bf R}=({\bf r},z)\), and \({\bf r}\) an in-plane positional vector rescaled from the 
original one according to the in-plane elastic constants \cite{Mikheev,Balents}. 
As will be revealed later, this case occurs when the mangeic field and the anisotropy parameter
are large enough, such that every block layer is occupied by Josephson vortices and the two unit  
vectors in real space are \({\bf a}_{1}=(d/f,0,0)\) and \({\bf a}_{2}=(d/2f,0,d)\).
The primitive reciprocal lattice vectors are \({\bf K}=(\pm 2f\pi/d,0,\pm \pi/d) \) and
\((0,0,\pm 2\pi/d)\).  The last two ones are equivalent to the origin.  In this case, dominant
thermal fluctuations are associated with the other four reciprocal vectors.  Especially, 
the \(k_c\) and \(k_x\) profiles of the partial structure factor with \(y=0\) around them are evaluated
as

\begin{equation}
\begin{array}{rl}
  S(\pm 2f\pi/d, k_c ) & \sim \int dxdz e^{ -i (k_c \pm\pi/d) z}  g( {\bf K}, {\bf R}) \\[10pt]
                         & \sim \frac{\xi_c^{-1}}{\xi^{-2}_c+(k_c\pm\pi/d)^2};  
\end{array}
\label{eqn:Lorentzian}
\end{equation}

\begin{equation}
\begin{array}{rl}
  S( k_x,\pm\pi/d ) & \sim \int dxdz e^{ -i (k_x\pm2f\pi/d) x}  g( {\bf K}, {\bf R}) \\[10pt]
                     & \sim |k_x\pm 2f\pi/d|^{-1+\eta},
\end{array}
\label{eqn:powerlaw}
\end{equation}

\noindent with \({\bf R}=(x,0,z)\), up to multiplicative coefficients and linear corrections.
Namely, in such a phase of 2D QLRO the structure factor should show power-law singularities 
in the \(k_x\) direction at reciprocal-lattice vectors, while smears out in the \(k_c\) direction in a
Lorentzian form.   By fitting the profiles of the Bragg spots, the correlation length
\(\xi_c\) and the exponent \(\eta\) of QLRO in \(ab\) planes can be evaluated.  
Detailed discussions on profiles of Bragg peaks in the smectic phase of liquid crystals
can be found in Ref.\cite{Als-Nielsen}.

\subsection{First-order melting transition: \(\gamma=8\)}

Let us start with a system of the anisotropy parameter \(\gamma=8\), which models the high-\(T_c\) superconductor
YBa\(_2\)Cu\(_3\)O\(_{7-\delta}\).   Upon temperature sweeping, a first-order transition is observed, 
indicated by the \(\delta\)-function peak in the specific heat and 
a kink anomaly in the Josephson energy at \(T_m=0.96J/k_B\), as displayed in Figs.1 and 4
of Ref.\cite{Huab}.    

\begin{figure}
\vspace{8.5cm}
\includegraphics{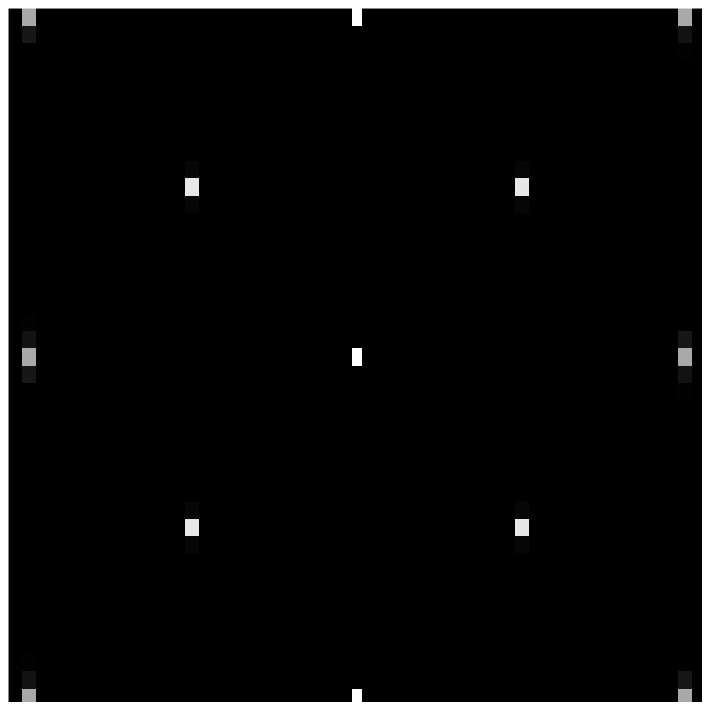}
\includegraphics{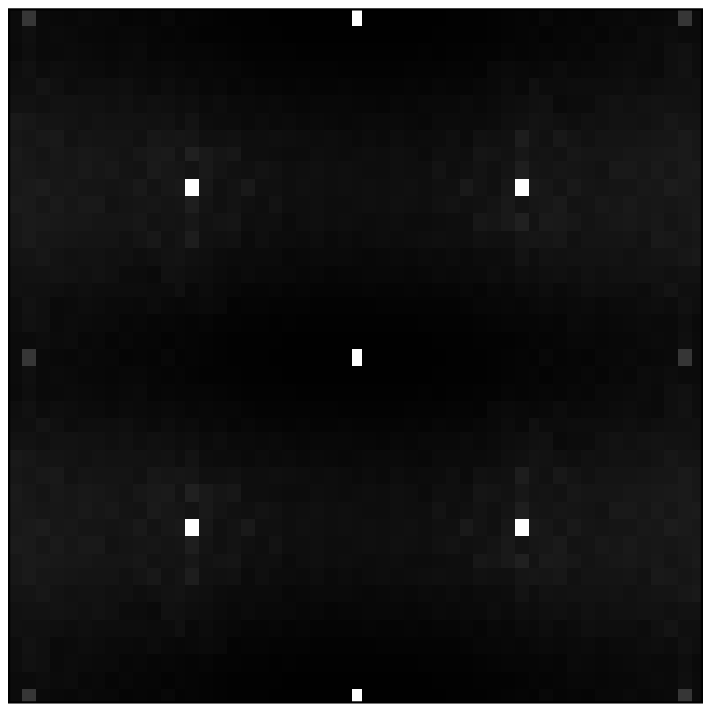}
\includegraphics{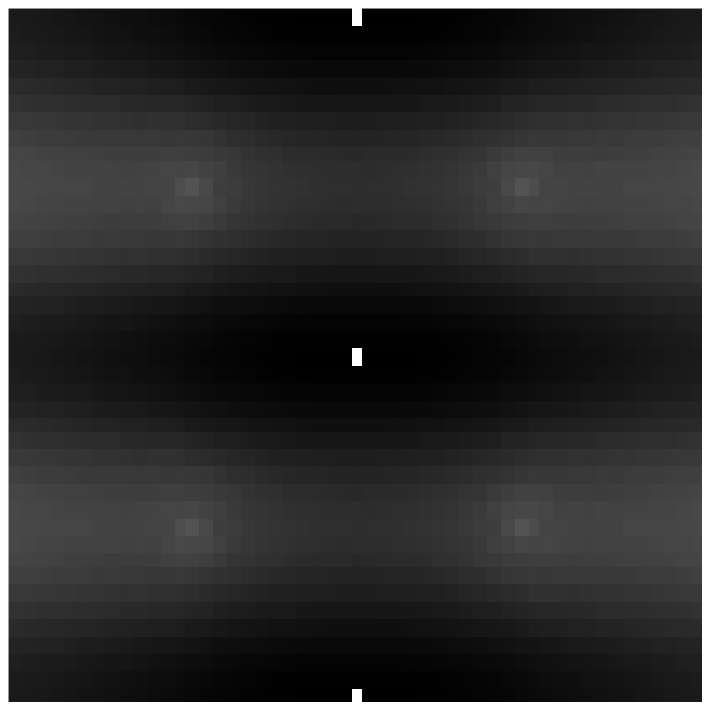}
\includegraphics{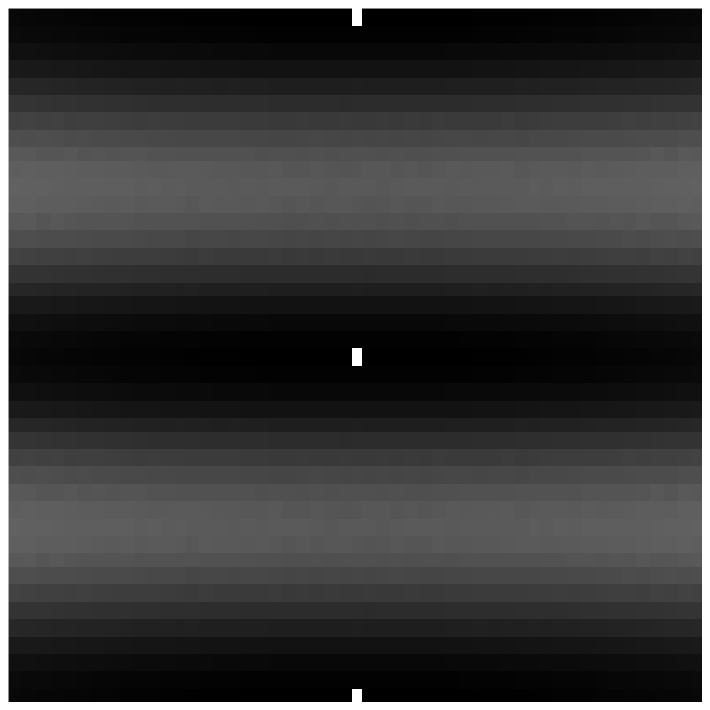}
\caption{\label{fig:S(q)g8}
Structure factors \(S(k_x,k_c)\) for \(\gamma=8\) at several typical temperatures.  
Top-left: \(T=0.1J/k_B\);
Top-right: \(T=0.96J/k_B\); Bottom-left: \(T=0.98J/k_B\); Bottom-right: \(T=1.5J/k_B\).
The panels are for wave numbers within 
\(k_x\in [-25\pi/192d, 25\pi/192d] \) (horizontal) and \(k_c\in [-2\pi/d, 2\pi/d] \) (vertical).
The spots at \((k_x,k_c)=(0,\pm 2\pi/d)\) are quivalent to \((0,0)\) in the present system. 
}
\end{figure}

\begin{figure}
\vspace{4cm}
\includegraphics{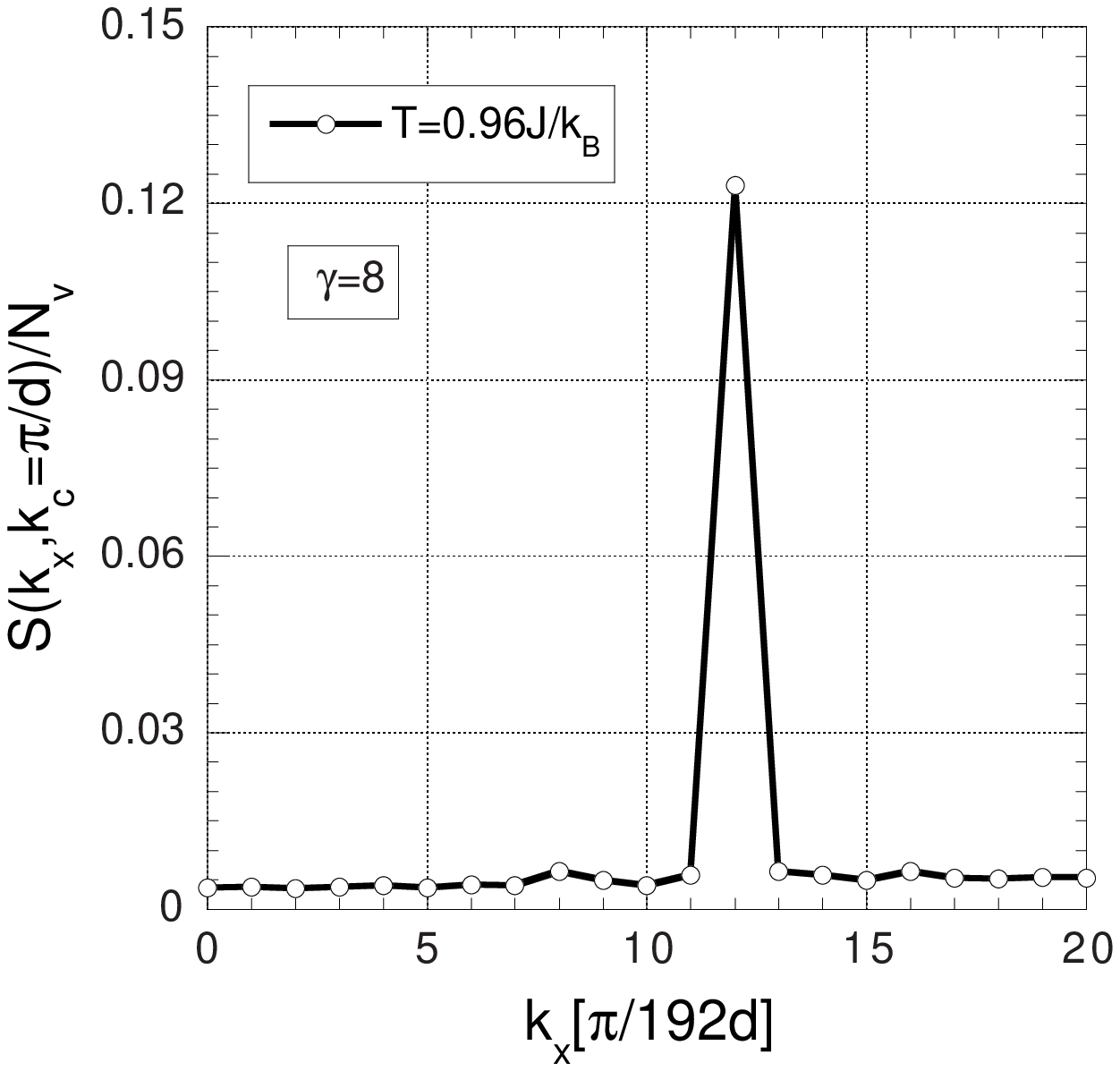}
\includegraphics{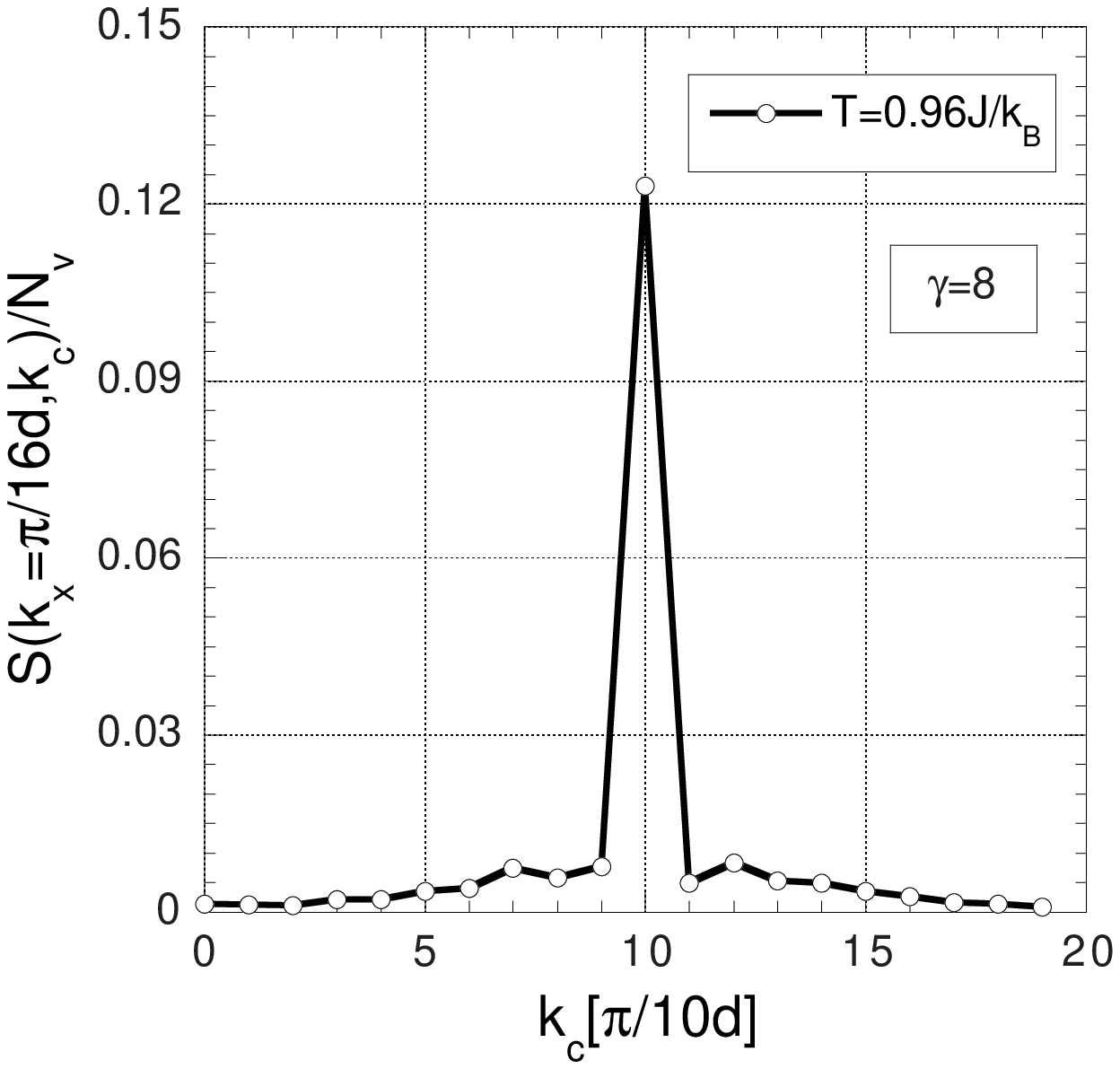}
\caption{\label{fig:profilesg8}
\(k_x\) (left) and \(k_c\) (right) profiles of Bragg peaks in the top-right panel of 
Fig.\ref{fig:S(q)g8} for \(\gamma=8\) at \(T=0.96J/k_B\).  
}
\end{figure}

In order to characterize the phase transition better,
we investigate the crystalline order of Josephson vortices in terms of the 
structure factors.  At low temperatures, there is a fine 3D LRCO in the
system indicated by the \(\delta\)-function Bragg peaks at the reciprocal-lattice
vectors, not only the minimal ones but also at higher-order satellites, as displayed in the
top-left panel of Fig.\ref{fig:S(q)g8}.
The Bragg peaks are observed up to \(T_m\simeq 0.96J/k_B\), without losing their sharpness
as shown in the top-right panel of Fig.\ref{fig:S(q)g8} and in Fig.\ref{fig:profilesg8}.
The 3D LRCO disappears when temperature crosses \(T_m\). The first-order transition at \(T_m\) 
is therefore identified as the melting of the Josephson vortex lattice, similar to the melting of
Abrikosov (or pancake) vortex lattice in magnetic fields parallel to the \(c\) axis.

At \(T\simeq 0.98J/k_B\), SR correlations are observed as shown in the bottom-left
panel of Fig.\ref{fig:S(q)g8}, which characterize a vortex liquid with layer modulation.  
It is worthy to
notice that, in spite of the high anisotropy in the vortex lattice, thermal fluctuations 
of large wave-lengths are quite isotropic for \(\gamma=8\) as reflected in the spots 
at \((k_x,k_c)=(\pm 2f\pi /d, \pm \pi /d)\) in the bottom-left panel of Fig.\ref{fig:S(q)g8}.  
At higher temperatures, \(T\ge 1.2J/k_B\), correlations are suppressed by thermal fluctuations 
to the scale of grid size of simulation as depicted in the bottom-right
panel of Fig.\ref{fig:S(q)g8}.  Containing huge number of closed vortex loops the system
is completely disordered.  In such a case, a coarse-graining in the \(ab\) planes
results in zero amplitude of the superconductivity order parameter, corresponding to the normal 
states.  In this temperature regime vortices are therefore not suitable for describing the system anymore.
The transformation from the vortex liquid to the normal phase
is a crossover, characterized by the broad peak in the specific heat in Fig. 3 of Ref.\cite{Huab}.

We have also simulated systems with \(1\le\gamma < 8\) at \(f=1/32\).  The melting point increases
with decreasing anisotropy parameter.  Josephson vortices distribute in a subset of
block layers for \(\gamma<5\) \cite{Huab}.  The phase transitions are first order at which
3D LRCO sets up for all the cases we have searched.  These simulation results look  
inconsistent with scenarios of continuous melting at intermediate and low magnetic fields.

\subsection{Multicritical point}

The first-order melting of Josephson vortex lattice is suppressed to continuous transitions
when the anisotropy parameter increases to \(\gamma=10\), when the filling factor is fixed
at \(f=1/32\), as reported in Ref. \cite{Huab}.    A critical value of the product between
the anisotropy parameter and filling factor may be given by \cite{Blatter,Huab}:   

\begin{equation}
f\gamma=\frac{1}{2\sqrt{3}}.
\end{equation}

It is easy to see that below the critical value, the Josephson vortex system
is physically equivalent to a 3D anisotropic Abrikosov vortex lattice.  Rescaling the system
by the anisotropy parameter should result in an effective isotropic system.  All physical 
properties, including possible phases and the nature of phase transition, should be essentially 
the same as  the isotropic system except for an anisotropy scaling
\cite{Blatter3}.  This physical discussion is consistent with our observations presented in the 
preceding subsection.  

On the other hand, the system becomes un-rescalable above the 
critical value,  since the minimal inter-vortex separation in the \(c\) axis \(2d\) has been reached
at the critical value.  Any increase in either \(f\) or \(\gamma\) results in overwhelming 
\(x\)-direction inter-vortex repulsions, which enhance the in-plane crystalline order. 
Although the CuO\(_2\) layers fix the inter-vortex distance for Josephson vortices in the \(c\) axis,
it has nothing to do with the inter-vortex correlations.  This is the most important peculiarity of
the interlayer Josephson vortices compared with liquid crystals in which smectic orders are realized.

\subsection{Intermediate phase: \(\gamma=20\)} 

\begin{figure}
\vspace{8.5cm}
\includegraphics{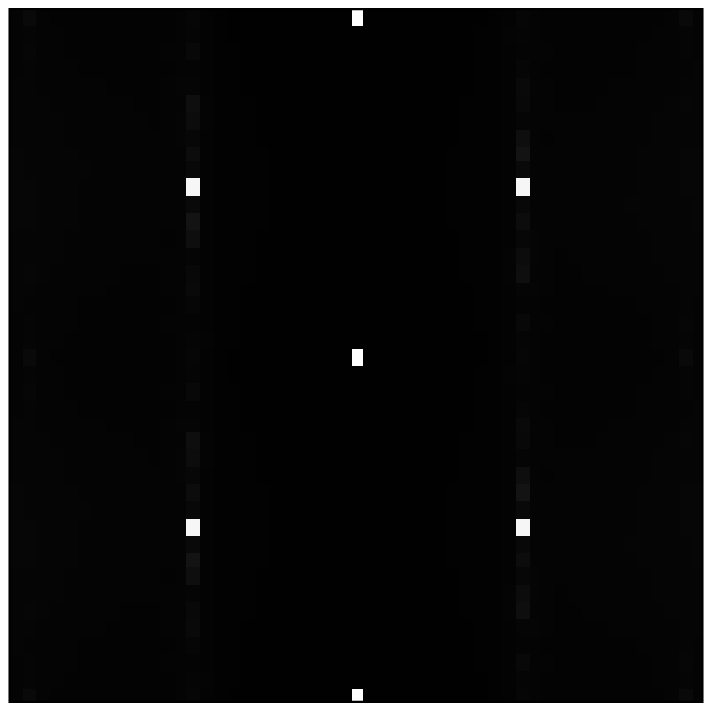}
\includegraphics{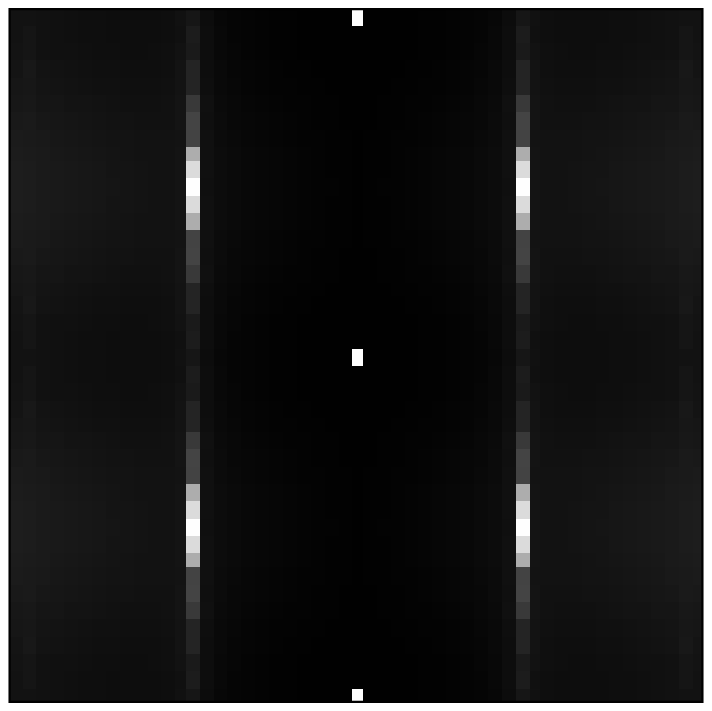}
\includegraphics{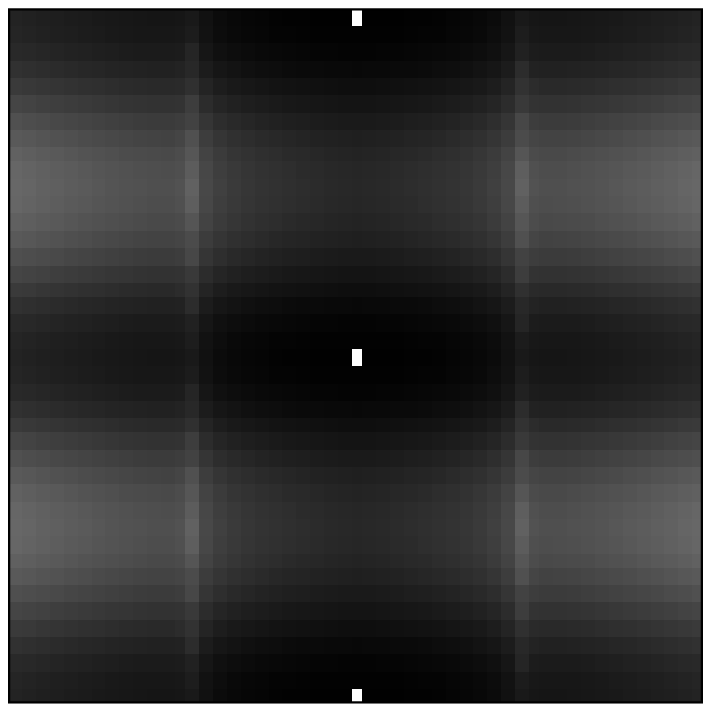}
\includegraphics{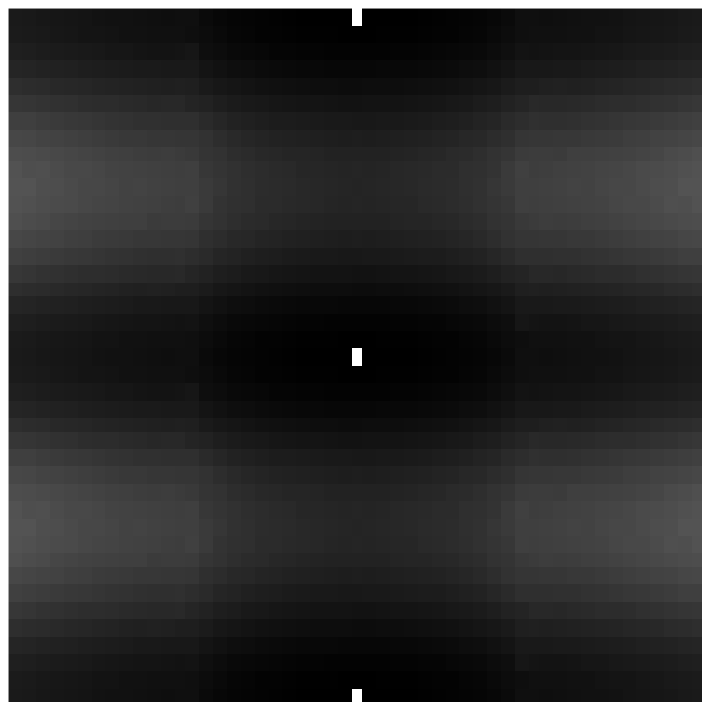}
\caption{\label{fig:S(q)g20}
Structure factors \(S(k_x,k_c)\) for \(\gamma=20\) at several typical temperatures.  Top-left: \(T=0.65J/k_B\);
Top-right: \(T=0.7J/k_B\); Bottom-left: \(T=0.95J/k_B\); Bottom-right: \(T=0.97J/k_B\).
The panels are for wave numbers within 
\(k_x\in [-25\pi/192d, 25\pi/192d] \) (horizontal) and \(k_c\in [-2\pi/d, 2\pi/d] \) (vertical). 
The spots at \((k_x,k_c)=(0,\pm 2\pi/d)\) are quivalent to \((0,0)\) in the present system. 
}
\end{figure}

\begin{figure}
\vspace{4cm}
\includegraphics{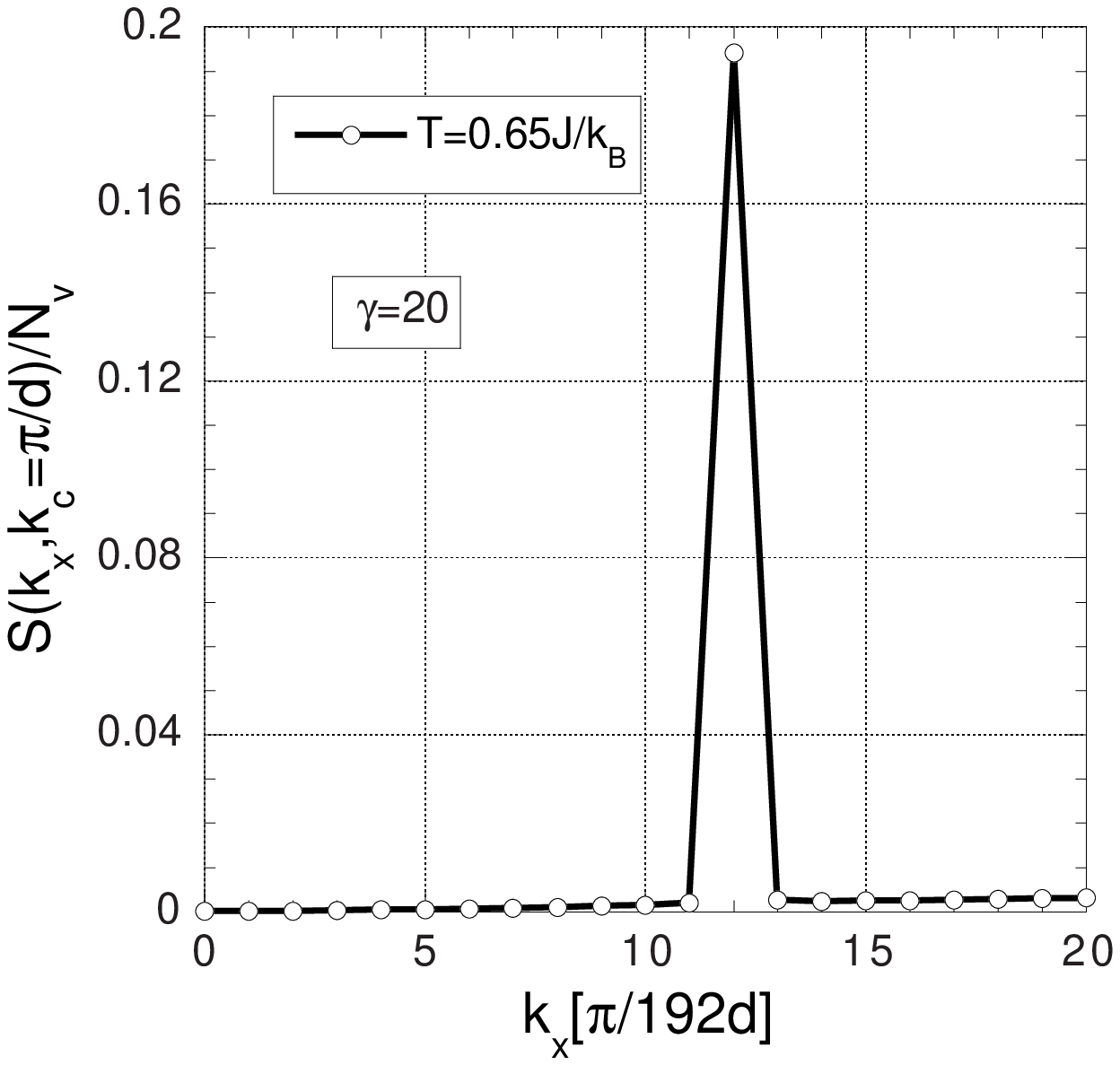}
\includegraphics{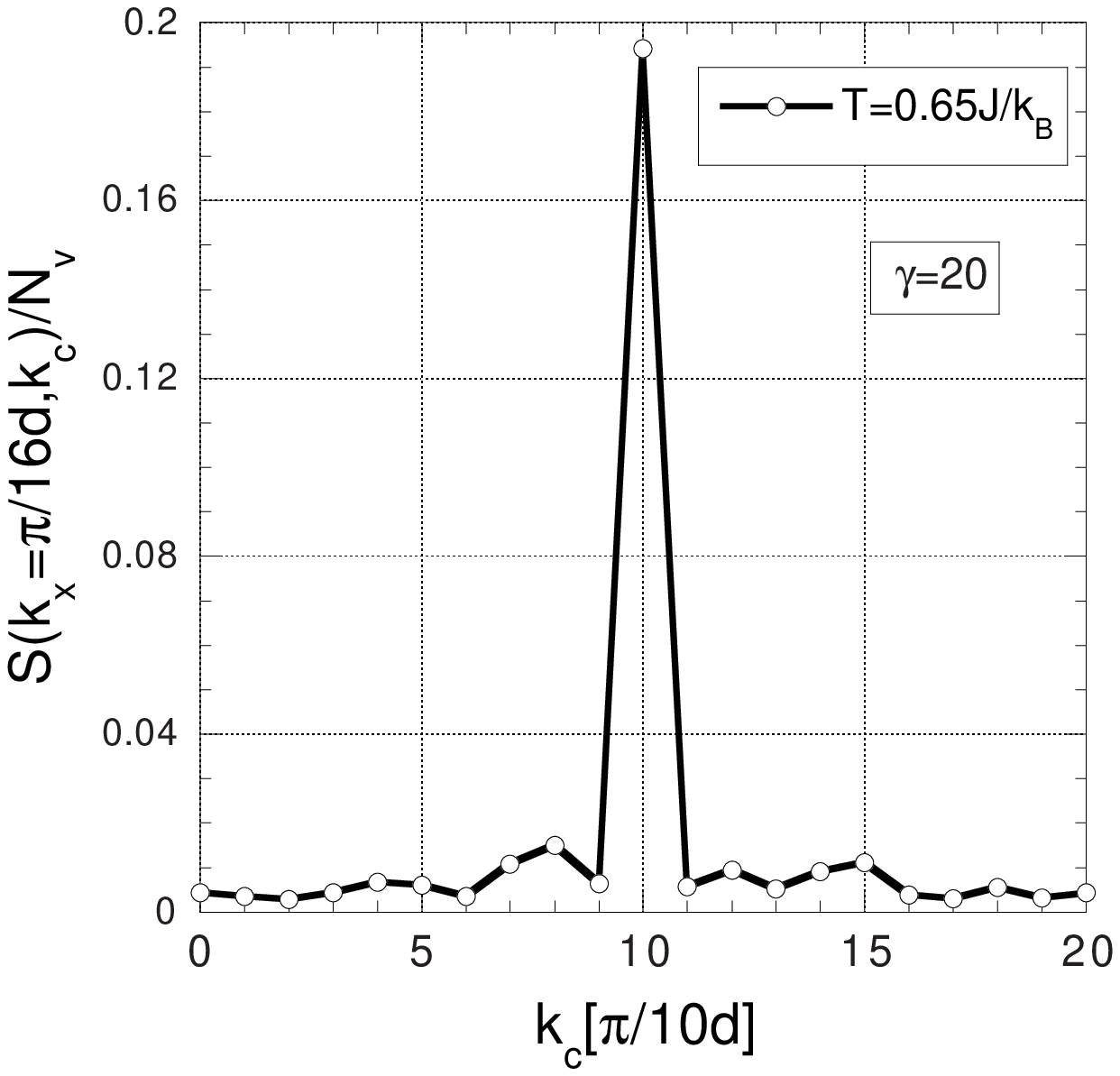}
\caption{\label{fig:profilest065}
\(k_x\) (left) and \(k_c\) (right) profiles of Bragg peaks in the top-left panel of 
Fig.\ref{fig:S(q)g20} for \(\gamma=20\) at \(T=0.65J/k_B\).  
}
\end{figure}

In order to reveal the crystalline order of Josephson vortices above the critical point, we choose 
\(\gamma=20\) which is quite above the multicritical value \(\gamma=16/\sqrt{3}\simeq 9.24\)
at \(f=1/32\).  An appropriately large anisotropy parameter realizes the intermediate phase in 
a wide temperature regime while leaves a reasonable interlayer Josephson coupling, and thus is 
convenient for computer simulations.   For \(T\le 0.65J/k_B\), a 3D 
LRCO same as that for \(\gamma=8\) is observed in this highly anisotropic case as shown in the 
top-left panel of Fig.\ref{fig:S(q)g20} and in Fig.\ref{fig:profilest065}.

In a sharp contrast to the case of \(\gamma=8\), however, the Bragg spots  for \(\gamma=20\) at 
\((k_x,k_c)=(\pm 2f\pi/d,\pm \pi/d)\) smear significantly in the \(k_c\) 
direction as temperature increases to \(T= 0.7J/k_B\), while leaving the 
sharpness in the \(k_x\) direction almost unchanged as seen in the top-right panel of Fig.\ref{fig:S(q)g20}.
Thermal fluctuations are therefore very anisotropic as well as the structure of the vortex lattice.  The diffusive 
and stripe-like Bragg spots survive to \(T\simeq 0.95J/k_B\) (see the bottom-left panel of Fig.\ref{fig:S(q)g20}).

\begin{figure}
\vspace{4cm}
\includegraphics{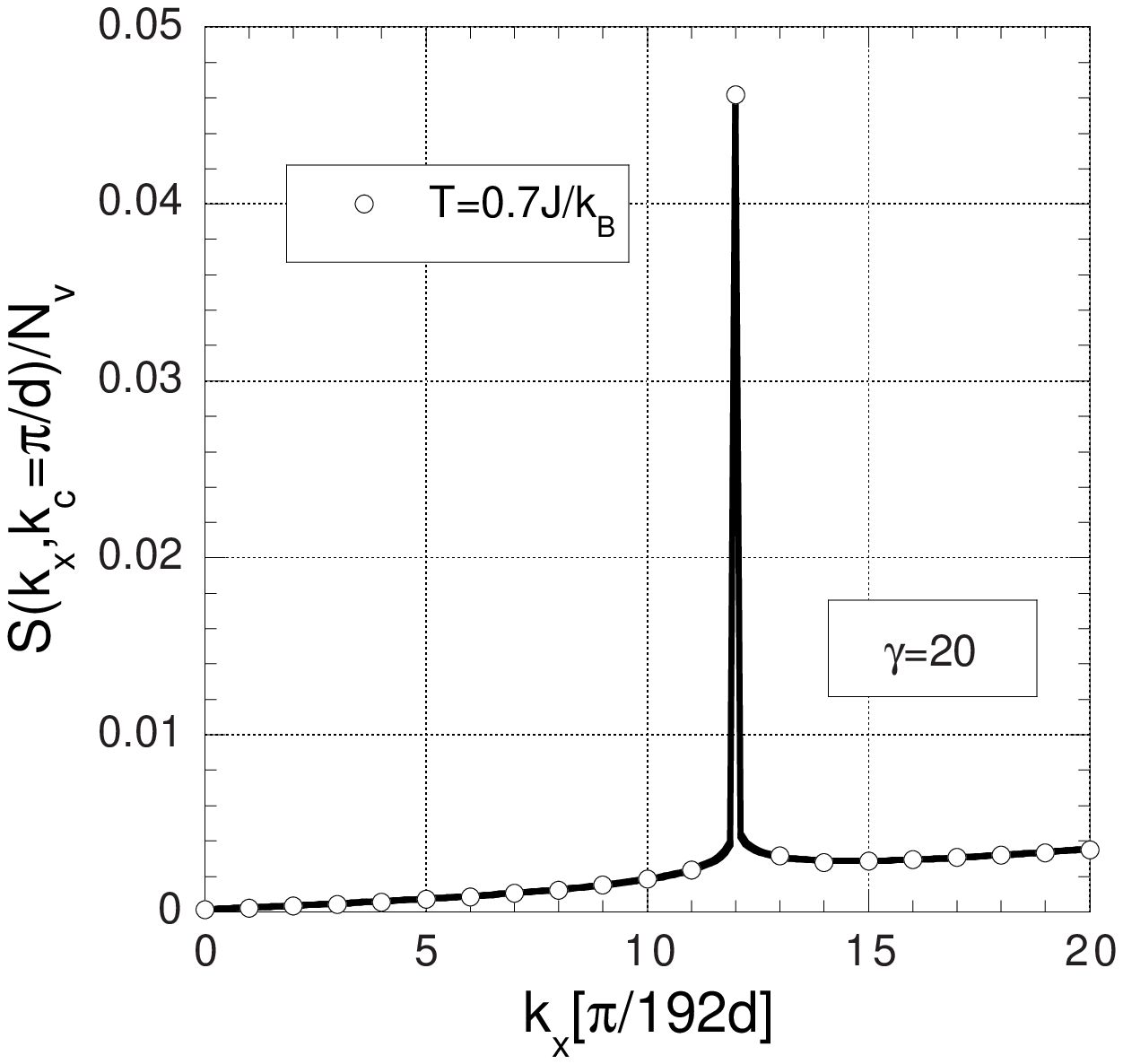}
\includegraphics{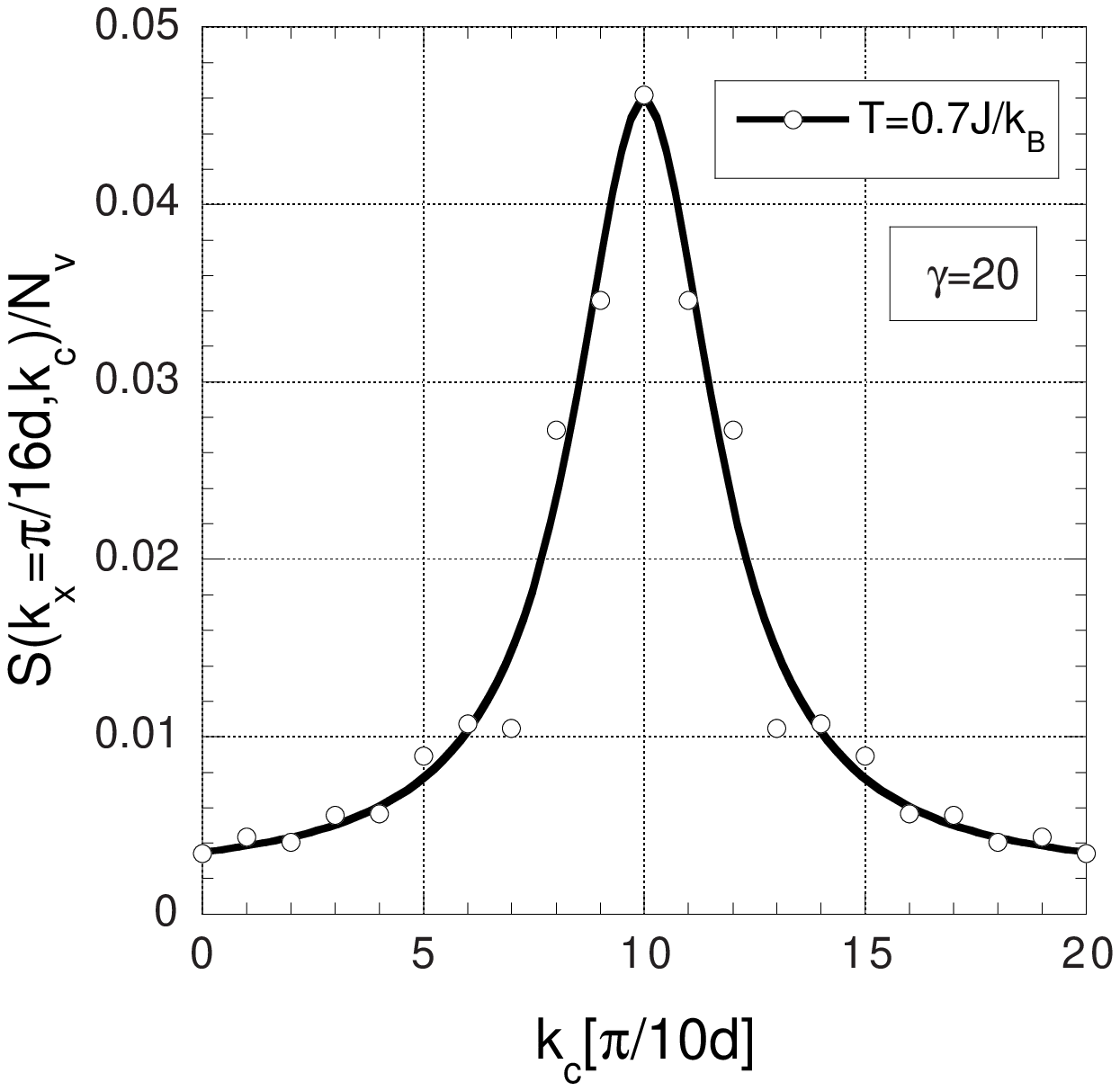}
\caption{\label{fig:profilest07}
\(k_x\) (left) and \(k_c\) (right) profiles of Bragg peaks in the top-right panel of 
Fig.\ref{fig:S(q)g20} for \(\gamma=20\) at \(T=0.7J/k_B\).
}
\end{figure}

As shown in the right panel of  Fig.\ref{fig:profilest07}, the \(k_c\) profile of the Bragg spots 
at \(T=0.7J/k_B\) is fitted well by the Lorentzian function in Eq.(\ref{eqn:Lorentzian})  with 
the correlation length \(\xi_c\simeq 1.5d\).   Therefore, the Josephson vortices are coupled to
each other only SR in the \(c\) axis.   Although it is not easy to distinguish numerically the
\(k_x\) profile in the left panel of  Fig.\ref{fig:profilest07} from a \(\delta\)-function Bragg peak,
it is easy to see that a \(\delta\)-function one is impossible as far as the system is 
decoupled in the \(c\) axis.  

\begin{figure}
\includegraphics[width=8cm,clip]{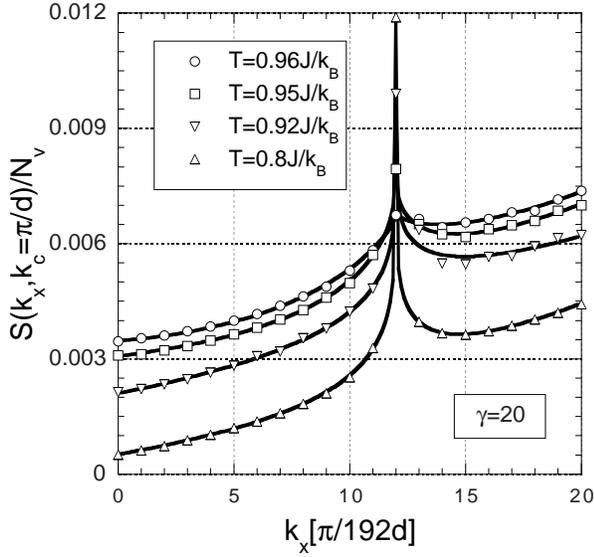}
\caption{\label{fig:S(qx)g20ts} 
\(k_x\) profiles of structrure factors \(S(k_x,k_c)\) around the Bragg spots for \(\gamma=20\)
in the intermediate phase. The solid curves are results 
of the least-squares fittings to the power-law function as described in text. 
} \end{figure}

The \(k_x\) profiles of structure factors \(S(k_x,k_c)\) for higher temperatures are depicted in 
Fig.\ref{fig:S(qx)g20ts}.  They are clearly different from the \(\delta\)-function peak for
\(T\le 0.65J/k_B\) for the present system, and those for \(\gamma=8\) in the whole
temperature regime \(T\le T_m\).  Supposing a cusp singularity in the profiles in
Fig.\ref{fig:S(qx)g20ts}, we can evaluate the exponent \(\eta\) in the correlation function
Eq.(\ref{eqn:rscf}) by fitting the data points including the peak values.  
This analysis results in \(\eta\simeq 1.97\pm 0.07, 1.42\pm 0.01, 
1.17\pm 0.007, 1.11\pm 0.003\) at \(T=0.96, 0.95, 0.92, 0.8J/k_B\), with 
the error bars from the least-squares fittings.  The fitting curves are shown in 
Fig.\ref{fig:S(qx)g20ts} by the solid curves.   The good quality of the fittings with \(\eta>1\) 
for all the temperatures,  especially for \(0\le k_x \le \pi/16d\), 
 can be taken as a justification of the assumption on the cusp singularity. 

\begin{figure}
\includegraphics[width=8cm,clip]{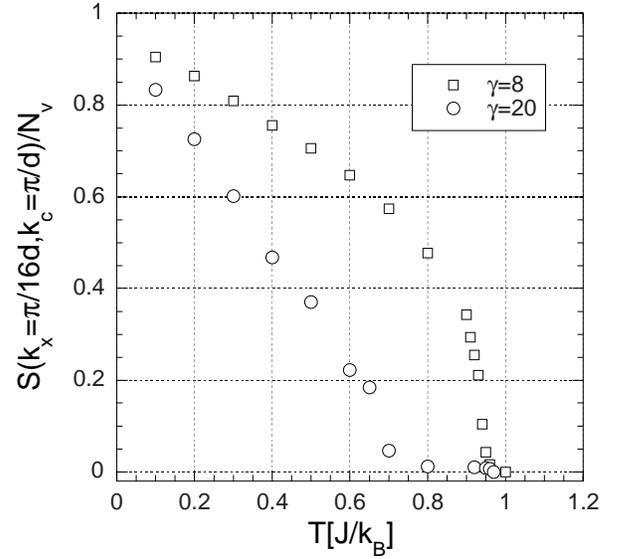}
\caption{\label{fig:Smax-T} 
Temperature dependence of the intensity of Bragg peaks for \(\gamma=8\) and 20.
}
\end{figure}

The temperature dependence of the maximal value of Bragg peaks is shown in Fig.\ref{fig:Smax-T}.
The intensity of Bragg spots is small and increases very slowly with decreasing temperature 
for \(0.7J/k_B\le T\le 0.95J/k_B\).  Across \(T_{\times}\) where the spots become \(\delta\)-function 
like, the intensity increases noticably. As a comparison, that for \(\gamma=8\) 
increases very sharply with decreasing temperature as soon as they become observable at
the melting point \(T_m\) as shown in Fig.\ref{fig:Smax-T}.  

One can also evaluate the partial structure factor \( S(k_x, k_y) \equiv S(k_x, k_y, z=0) \),
which describes the Josephson vortex correlations in the same block layers.
The profiles around the Bragg spots \((k_x,k_y)=(\pm 2f\pi/d,0)\) should behave as

\begin{equation}
S(k_x, 0) \sim |k_x\pm 2f\pi/d|^{-2+\eta}, \hspace{3mm} S(2f\pi/d, k_y)\sim |k_y|^{-2+\eta}.
\end{equation}

\begin{figure}
\includegraphics[width=8cm,clip]{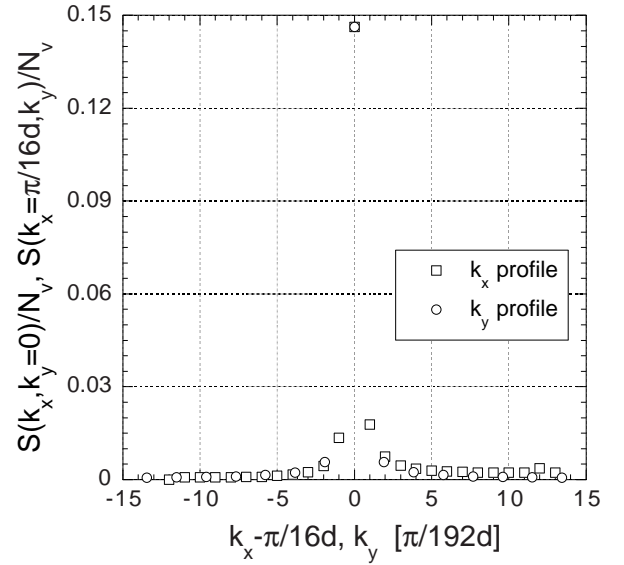}
\caption{\label{fig:Sqxqyg20t08} 
Profiles of the structure factor \(S(k_x,k_y)\) around the Bragg spot \((k_x,k_y)=(\pi/16d,0)\)
for \(\gamma=20\) at \(T=0.8J/k_B\).
}
\end{figure}

\noindent The simulated results are displayed in Fig.\ref{fig:Sqxqyg20t08} for \(T=0.8J/k_B\).
A power-law singularity is observed, which is consistent with
that in the \(x-c\) crosssections.  The collapse of two profiles left to the peak in Fig.\ref{fig:Sqxqyg20t08}
indicates that the in-plane correlation functions are governed by the same
exponent \(\eta\).   The deviation on the right side of the peak is obviously caused by the
second peak in the \(k_x\) profile at \(k_x=4f\pi/d\), which is absent in the \(k_y\) direction.  
Since \(\eta<2\) observed in the structure factor \(S(k_x,k_c)\), the unnormalized profiles 
should diverge in the thermodynamic limit as in the above equations.   Therefore, one cannot evaluate
the exponent \(\eta\) directly from the profiles.

The above observations therefore indicate the existence of an intermediate phase for \(\gamma=20\),
characterized by SR interlayer correlation and in-plane QLRO, 
which is clearly absent for \(\gamma=8\) below the multicritical point.

Is an intermediate phase plausible in the present system theoretically?  According to the
perturbative RG expansion to a 2D system by Balents and Nelson \cite{Balents}, the lattice is stable at 
low temperatures 
with \(\eta<1/2\) where dislocations caused by hops of segments of Josephson flux lines across the 
superconducting layers are irrelevant while the interlayer coupling is relevant; 
at high temperatures with \(\eta>2\) where the dislocations are relevant while the interlayer
coupling is irrelevant, the system behaves as liquid.  In the intermediate temperature regime,
both of them are relevant and the fate of the system is not very well controlled.  The
system may take one of the two states, or it can take the smectic order as the authors suggested.

\begin{figure}
\includegraphics[width=8cm,clip]{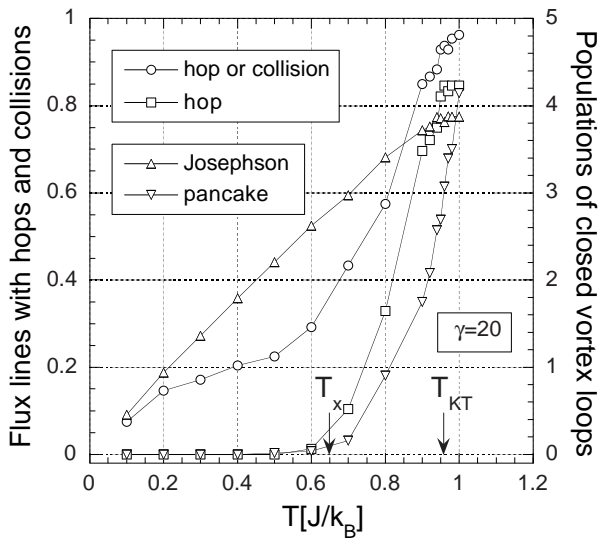}
\caption{\label{fig:hops-T} 
Temperature dependence of ratios of Josephson flux lines which 
contain segments hopping into neighboring block layers, of those which either hop or
collide with neighbors in the same block layers, and of populations of thermally excited, 
closed loops of Josephon vortices (normalized by \(40\times 240\)) and of those
containing also pancake vortices (normalized by 240).
}
 \end{figure}

In simulations, we analyze the trajectories of the Josephson flux lines when they travel through the sample. 
As depicted in Fig.\ref{fig:hops-T}, Josephson flux-lines are completely confined by the CuO\(_2\) layers 
below \(T_{\times}\).  Without dislocations caused by the hops, the system is in lattice phase as
shown in the top-left panel of Fig.\ref{fig:S(q)g20}, consistent with the theory.  
Dislocations become popular in the system when temperature is elavated cross \(T_{\times}\) as shown in 
Fig.\ref{fig:hops-T}.   It is interesting to observe that the maximal value of the exponent \(\eta\) in our 
simulations, \(\eta\simeq 1.97\) at \(T\simeq 0.96J/k_B\), is close
to the theoretical prediction \(\eta_c=2\) for the phase boundary between liquid and the intermediate phase.  
However,  we should notice that while the theory predicted relevance of interlayer coupling in the 
intermediate regime, our simulations indicate the SR \(c\)-axis correlation.  The origin of this discrepancy is not
clear at this moment.  As pointed out in Ref.\cite{Mikheev}, one possibility is that although the long-wave-length
displacements of Josephson flux lines in the neighboring layers are coupled, accumulation of disorders in a
stack of layers leads to decoupling, namely SR \(c\)-axis correlation. 

In Fig.\ref{fig:hops-T}, we also display the temperature dependence of population of closed
Josephson vortex loops activated thermally.  These excitations have not been treated
in the elastic theory, and are responsible for the temperature dependence of the exponents
\(\eta\).  The elastic constants
in the present system are therefore not simply dominated by entropy as discussed in the
elastic theory.   Our system is similar to the thermotropic smectics in this aspect \cite{Als-Nielsen}.
A counter example is found in the lyotropic 
lamellar phase of a multilayer fluid membrane system \cite{Safinya}.   
See \cite{Lubensky} for a detailed discussion.

\subsection{Novel Kosterlitz-Thouless phase}

Since the density-density correlations are SR in the \(c\) direction in the intermediate phase,
the effective free energy is governed by the long-wave-length fluctuations given in 
the following density expression

\begin{equation}
\rho ( {\bf r} ) = \psi(  {\bf r} ) e ^{ -i {\bf k}_0 \cdot {\bf r} }  
                     + \psi^*(  {\bf r} ) e ^{ i {\bf k}_0 \cdot {\bf r} },
\end{equation}

\noindent with \({\bf k}_0\equiv (k_{x0},k_{y0})=(2f\pi/d,0)\) and 

\begin{equation}
\psi (  {\bf r} ) = |\psi| e ^{ i2f\pi u_x/d}.
\end{equation}

When a small amplitude \( |\psi|\) is set up \cite{Balents}, presumably at \(T\simeq 0.96J/k_B\) as
in Figs.\ref{fig:S(qx)g20ts} and \ref{fig:Smax-T}, the effective free energy is reduced to

\begin{equation}
 F=\int d^2r(\triangledown u_x)^2.
\end{equation}

\noindent  Since the displacement field \(u_x\) is continuous and of the
modulus \(d/f\), the above free energy is effectively the same as the
Hamiltonian of the 2D XY model.  The intermediate phase 
therefore falls onto the KT fixed line.

\section{Superconductivity transition}

So far we have concentrated on the crystalline order of Josephson vortices.
It is also important to investigate the superconductivity in the system.  Although
the amplitude of the {\it local} order parameter of superconductivity is fixed
in the Hamiltonian (1), the fate of macroscopic superconductivity 
is governed by the order of Josephson vortices.   We measure the LR
superconductivity by the helicity modulus \cite{Teitel,Huc}.

\begin{figure}
\includegraphics[width=8cm,clip]{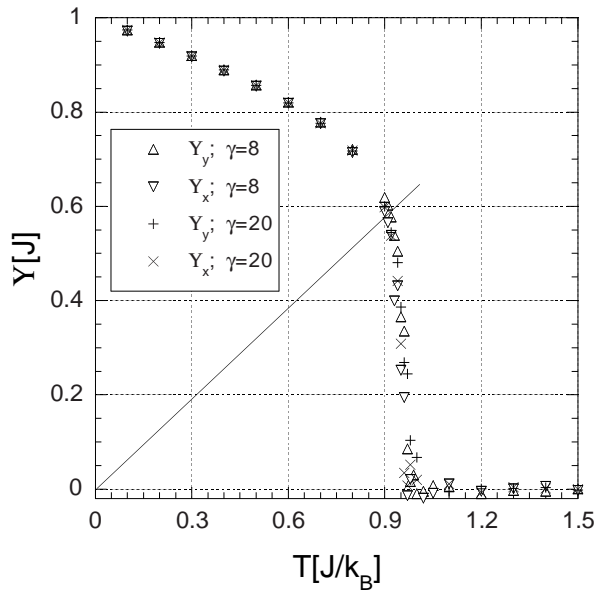}% Here is how to import EPS art
\caption{\label{fig:Upsilon} 
Temperature dependence of in-plane helicity moduli for \(\gamma=8\)
and \(\gamma=20\).  The solid line is for theoretical prediction of the 
universal jump at KT transitions.  For a pure 2D unfrustrated XY model corresponding
to \(\gamma=\infty\) in Hamiltonian (1) \(T_{\rm KT}^{\rm bare}\simeq 0.89J/k_B\).
}
\end{figure}

As shown in Fig.\ref{fig:Upsilon}, in-plane helicity moduli for \(\gamma=8\) set up 
sharply at the melting point \(T_m\simeq 0.96J/k_B\), where 3D LRCO
is realized.  The finite helicity modulus \(\Upsilon_y\) along the magnetic field
indicates the breaking of U(1) gauge symmetry, and thus the appearance of LRO of superconductivity.  
The finite \(\Upsilon_x\) is, on the other hand, the signature of the
intrinsic pinning on Josephson vortices from the layer structure, since the helicity modulus 
\(\Upsilon_x\) measures the energy cost for sliding Josephson vortices in the \(c\) 
direction.  \(\Upsilon_c\) is vanishing down to zero temperature, reflecting the absence
of pinning force in the \(x\) direction.  Therefore, the system is superconducting
in \(ab\) plane for \(T\le T_m\).  The sharp drop of the in-plane helicity moduli at the 
melting point is in accordance with the first-order nature of the melting transition.

\begin{figure}
\vspace{7.5cm}
\includegraphics{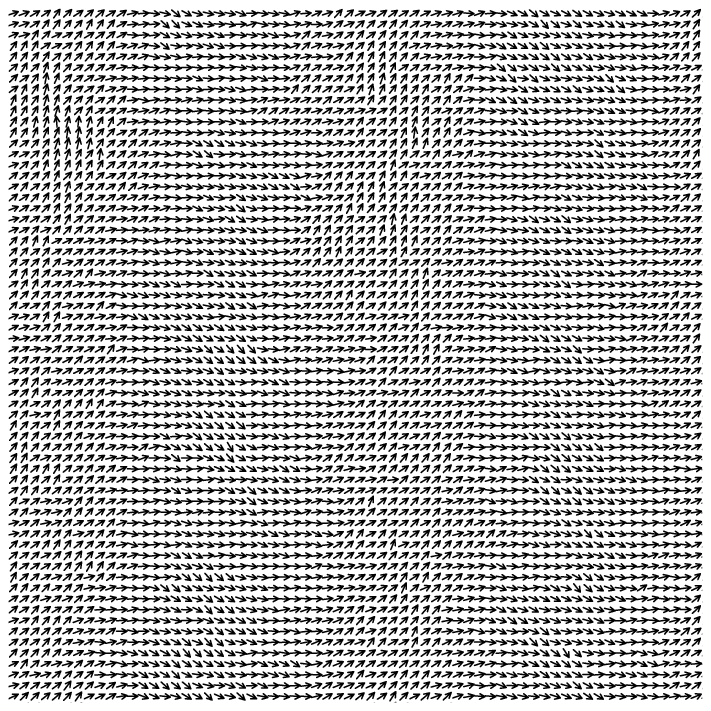}
\caption{\label{fig:phaseconfab10}
A typical configuration of phase variables on an \(ab\) plane at \(T=0.1J/k_B\) for 
\(\gamma=8\).  Shown is a region of dimensions \(l_x\times l_y=64d\times 64d\).
}
\end{figure}

A typical configuration of phase variables on an \(ab\) plane at low temperatures 
is displayed in Fig.\ref{fig:phaseconfab10}.   The phases are uniform 
along the magnetic field.  It is also regular in the \(x\) direction, except for 
a single-wave-number modulation governed by the density of Josephson flux lines
\cite{Bulaevskii,Korshunov}.  

\begin{figure}
\vspace{4cm}
\includegraphics{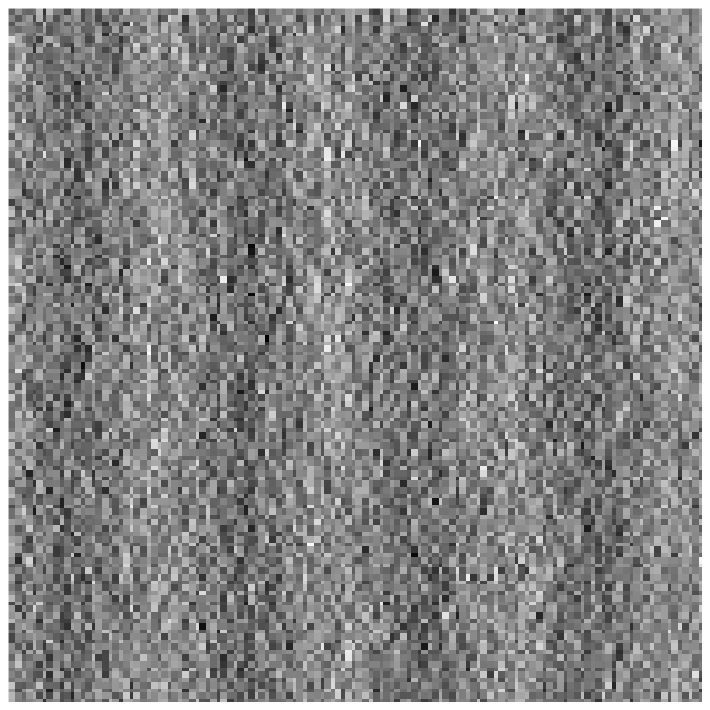}
\includegraphics{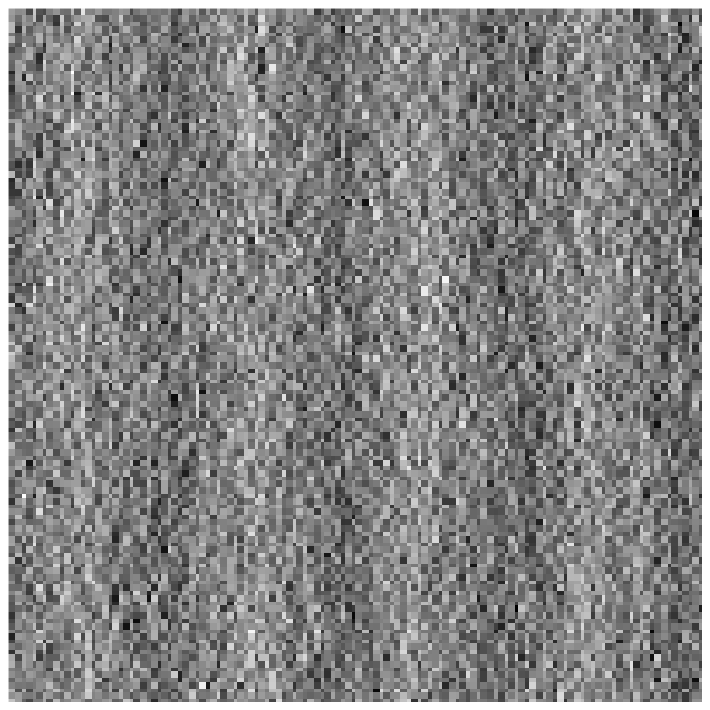}
\caption{\label{fig:phasewave}
Supercurrents between two sites neighboring to each other in the \(x\)
direction.  Shown is the region of the same origin of Fig.\ref{fig:phaseconfab10}
but of larger dimensions  \(l_x\times l_y=128d\times 128d\). 
}
\end{figure}

The static {\it phase wave} can be displayed in a more transparent way by showing 
the supercurrent in the \(x\) direction 
\(I_x({\bf r})\sim\sin(\varphi({\bf r}+\hat{\bf x}) - \varphi({\bf r}) )\). 
The supercurrents associated with the phase distribution in Fig.\ref{fig:phaseconfab10} 
are displayed in the left panel of Fig.\ref{fig:phasewave}.  Regions of
positive and negative supercurrents in the \(x\) direction are decorated by white
and black tones respectively.   There are four black (and white) stripes in the region of
\(l_x=128d\), corresponding to the density of Josephson flux lines \(f=1/32\).  The 
pattern of supercurrents on a neighboring \(ab\) plane is shown in the
right panel of  Fig.\ref{fig:phasewave}.  The stripes on the two planes
are opposite to each other in black and white tones, and sustain four 
Josephson flux lines between them shown in Fig.2 of Ref.\cite{Huab}.

\begin{figure}
\vspace{2.8cm}
\includegraphics{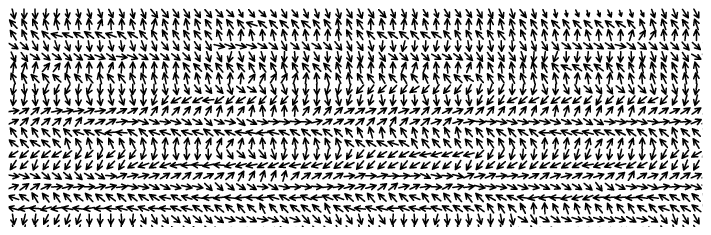}
\caption{\label{fig:phaseconfac1}
A typical configuration of phase variables on an \(ac\) section at low temperatures for 
\(\gamma=8\).  Shown is a region of dimensions \(l_x\times l_z=64d\times 20d\).
}
\end{figure}

Variations of the phase variables along the \(c\) direction is displayed in 
Fig.\ref{fig:phaseconfac1}.  The configuration is similar to the sequence 
(\(0, 0, \pi, \pi, 0, 0, \cdot\cdot\cdot\)) derived in Ref.\cite{Radzihovsky}.
The simulated \(c\)-axis modulation is more complex, which together with that
in the \(x\) direction minimizes the energy for weak but finite interlayer couplings.  

The amplitude of the phase modulation in \(x\) direction is small around and above 
the multicritical point in accordance with the analytic result by Bulaevsky and 
Clem: \({\cal A}=1/2\pi(\gamma f)^2<1\) \cite{Bulaevskii}.  Most of the frustrations
induced by the magnetic field are confined in the \(c\) direction as clearly seen in 
Fig.\ref{fig:phaseconfac1}.  The large phase modulations in the \(c\) direction make the phase slip hard, 
resulting in the finite helicity modulus \(\Upsilon_x\).  The small, and thus soft phase modulations 
in the \(x\) direction suppress the helicity modulus \(\Upsilon_c\) to zero.

For \(\gamma=20\), finite in-plane helicity moduli signal the onset 
of QLRO of superconductivity, as a twin of the QLRCO
of Josephson vortices. The sharp drops of helicity moduli for \(\gamma=20\) correspond
to the universal jump in KT transitions \cite{Nelson}.

The jump of helicity modulus in Fig.\ref{fig:Upsilon} is rounded from the universal one by the 
finite-size effect as usual.  
An analysis on finite-size effects is certainly helpful here.  However, as one can see in a 
careful study by Lee and Teitel on a 2D Coulomb gas system in Ref.\cite{Teitel2}, 
this approach is usually very hard since the finite-size effects are logarithmically weak
\cite{Minnhagen}.  A similar analysis for the
present system should be even harder, since first, there is a SR correlation in the
\(c\) direction, secondly, the system size should be very huge in order to contain
sufficient number of Josephson vortices,  and thirdly, the present system is anisotropic
in all three directions.  

For these reasons, we will not seek to verify the universal jump 
in the present system.  Instead, upon revealing the existence of the novel KT phase 
from simulation results and the symmetry argument in terms of the crystalline order 
of Josephson vortices in the preceding section, we adopt the expected universal
jump \cite{Nelson}

\begin{equation}
\Upsilon/k_BT_{\rm KT}=2/\pi.
\label{eqn:univjump}
\end{equation}

\noindent to estimate the KT 
transition temperature \cite{Minnhagen2}.  From data in Fig.\ref{fig:Upsilon}, we find
\(T_{\rm KT}\simeq 0.91J/k_B\). 

The KT transition temperature for \(\gamma=20\) is above that of the isolated 2D system
\(T_{\rm KT}^{\rm bare}\simeq 0.89J/k_B\).  This is physical since couplings between nearest 
neighboring layers enhance ordering.  The bare KT 
transition point sets a lower bound for the normal to 
superconductivity transition in magnetic fields parallel to the \(ab\) plane, as can be read 
from Hamiltonian (1).

An important feature in the helicity moduli for  \(\gamma=20\) is revealed in Fig.\ref{fig:Upsilon}.
Namely, the helicity moduli in \(x\) and \(y\) directions are collapsing at all temperatures.   
Presuming that the relation between the exponent \(a (T)\) defined in the 
I-V characteristics \(V\sim I^{a (T)}\) and the helicity moduli  
\cite{Ambegaokar,QHChen}

\begin{equation}
a_{\mu}=1+\Upsilon_{\mu}\pi/k_BT, \hspace{5mm} {\rm for} \hspace{5mm} \mu=x,y
\end{equation}

\noindent in the KT phase derived in pure 2D systems is also applicable for the present KT phase,
the collapse of helicity moduli in the two in-plane directions is consistent with   
the orientation-independent, {\it i.e.} Lorentz-force-independent, dissipations when 
the current and magnetic field are both parallel to the \(ab\) plane \cite{Iye}.
Further investigation is expected in order to address this point completely.
 
One might notice that in Fig.\ref{fig:Upsilon} the helicity moduli for the two anisotropy parameters
are very close to each other.  It is reasonable since, first, both first-order and KT transitions reveal
sharp drops in the helicity modulus, and secondly, the anisotropy parameters are both very large such
that the transition points are close to the lower bound \(T^{\rm bare}_{\rm KT}=0.89J/k_B\).
The helicity moduli in the \(x\) and \(y\) directions are close to each other for \(\gamma=8\) merely because
that it is close to the multicritical point.  For smaller anisotropy parameters,
thus far from the multicritical point, we find clearly that the isotropy in the in-plane helicity
modulus is broken.  This explains the anisotropic resistivity in samples of YBa\(_2\)Cu\(_3\)O\(_{7-\delta}\). 

\section{Phase transition at \(T_{\rm x}\)}

What is the nature of the phase transition between the 2D QLRO and
3D LRO at \(T_{\times}\)?   In order to answer this question, we follow
Balents and Nelson to compose an effective Landau free-energy functional for the
3D Josephson vortex lattice, which was formulated for a possible smectic to crystal
transition \cite{Balents}.  The vortex density in the 3D crystal phase is expressed 
by three primitive reciprocal-lattice vectors as in Eq.(\ref{eqn:density}).
A Landau free energy for the dominant fluctuations is given by, up to a spatial anisotropy in 
the coefficients which is unimportant here, 

\begin{equation}
\begin{array}{rl}
F & =1/2\int d^3R\Big[ h|\bigtriangledown\Psi_1|^2 + h|\bigtriangledown\Psi_2|^2 \\[10pt]
   &                         + g ( \bigtriangledown\Psi_1 \bigtriangledown\Psi_2
                                   + \bigtriangledown\Psi^*_1 \bigtriangledown\Psi^*_2 )  \\[10pt]
  &                        +r ( |\Psi_1|^2 +  |\Psi_2|^2 ) + w ( \Psi_1\Psi_2 +\Psi^*_1\Psi^*_2 )  
                            + \cdot\cdot\cdot \Big],
\end{array} 
\end{equation}

\noindent where the order parameter \(\Psi_0=1\) associated with \((k_x,k_c)=(0,\pm 2\pi/d)\) 
has been included into the coefficients
\(g\) and \(w\).  Our simulations have revealed that Josephson flux lines are completely confined 
by superconducting layers, namely  \(u_c=0\), for \(T<T_{\times}\).  Thus, the two order
parameters in Eq.(\ref{eqn:op}) become complex conjugate to each other:

\begin{equation}
\Psi_1=\Psi^*_2=|\Psi| e^{i2f\pi u_x/d}.
\end{equation}

\noindent The free energy is then reduced to

\begin{equation}
F=\int d^3R( \triangledown u_x )^2.
\end{equation}

\noindent The phase transition is therefore second order with the critical phenomena
governed by the universality class of the 3D XY model.

A phase transition in this universality class possesses a negative critical exponent
\(\alpha\) for the specific heat.  This naturally explains the smooth temperature
dependence of the simulated specific heat around \(T_{\times}\) in Fig. 3 of
Ref.\cite{Huab}.  

When the critical point \(T_{\times}\) is approached from below, interlayer shear 
modulus of the Josephson vortex lattice is softened continuously to zero as 
\(C_{66}\sim d/\xi_c \sim (T_{\times}-T)^{\nu} \) with \(\nu\simeq 0.67\), and
ramains vanishing in the whole intermediate phase.

\section{Phase diagram}

\begin{figure}
\vspace{6.5cm}
\includegraphics{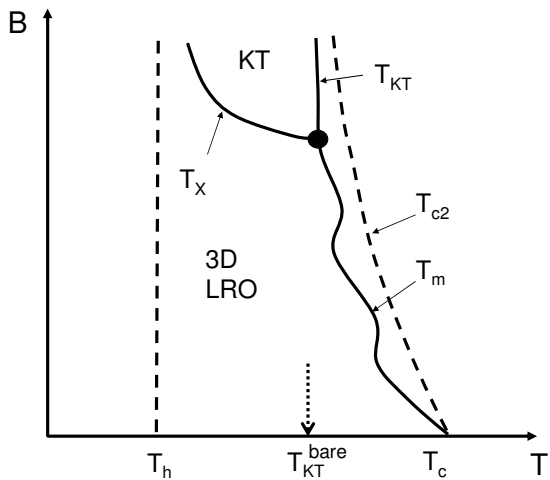}
\caption{\label{fig:pd}
\(B-T\) phase diagram for interlayer Josephson vortices with a multicritical point at
\(B_{\rm mc}=\phi_0/2\sqrt{3}\gamma d^2\).
The phase boundaries \(T_m(B), T_{\rm KT}(B)\) and \(T_{\times}(B)\) 
are associated with first-order, KT and 3D XY phase transitions as discussed in text. 
Possible crossovers are included by dashed lines.
}
\end{figure}

Based on the analyses presented so long, we map out in Fig.\ref{fig:pd} the \(B-T\) phase diagram 
for interlayer Josephson vortices, noting that the same physics should occur 
when the magnetic field is tuned at a given anisotropy parameter. The first-order 
melting line at low magnetic fields branches into two phase boundaries at the multicritical point 
characterized by the magnetic field

\begin{equation}
B_{mc}=\frac{\phi_0}{2\sqrt{3}\gamma d^2}.
\end{equation}

\noindent The phase at high magnetic fields and intermediate temperatures is a 
novel KT phase.

There are two dashed, crossover curves in the phase diagram Fig.\ref{fig:pd}. The high-temperature one
denoted by \(T_{c2}\) follows the broad cusp in the specific heat where huge numbers of vortex loops
are excited.  It is a crossover where the amplitude of superconductivity order parameter attains a finite 
value, and thus corresponds to \(H_{c2}\) in the mean-field theory.
The low-temperature and almost-field-independent line  \(T_h\) denotes a crossover
temperature below which thermally activated pancake vortices are very few and bound to each other 
too tightly such that Josephson flux lines cannot hops across the 
CuO\(_2\) layers.  Transport properties can be different on the two sides of the crossover line \(T_h\)
since randomly distributed point-like defects might pin Josephson flux lines via
the pancake vortices present above \(T_h\) \cite{Hirata,Zhukov}.  As the displacement in the \(c\) direction
is massive below \(T_h\), one expects the transverse Meissner effect in the low temperature regime.

\section{Summary and Discussions}

The main finding of the present work is that the normal to superconductivity phase transition in
magnetic fields parallel to superconducting layers is first order at low fields, while two-step and
continuous with a KT type intermediate phase at high fields.  There is a multicritical point at
magnetic field of order \(B_{\rm mc}= \phi_0/2\sqrt{3}\gamma d^2\) in the \(B-T\) phase diagram.

That the above phases are genuine for the interlayer Josephson
vortices rather than artifacts from finite-size and short-time effects can be concluded from 
the following considerations:  First,
the successful observation of the single, first-order melting transition for \(\gamma=8\)
indicates that the system size is sufficient for the onset of 3D LRCO
below the multicritical anisotropy parameter;  Second, the SR order in the \(c\) direction for
\(\gamma=20\) is not a finite-size effect, since periodic boundary conditions are adopted in the present
study, which tend to enhance ordering.  It is not a short simulation-time effect either, since a heating process
from a 3D LR crystalline order is adopted;  Third, it is clear that the system described by Hamiltonian (1) 
with a finite \(\gamma\)
should order (maybe partially) at a temperature higher than that of the limit case \(\gamma=\infty\), which 
composes of independent 2D superconductors with \(T_{\rm KT}^{\rm bare} \simeq 0.89J/k_B\).  Thus, the
in-plane vortex correlation functions at, for example, \(T=0.7\) and \(0.8J/k_B<T_{\rm KT}^{\rm bare}\) for 
\(\gamma=20\) cannot be SR; with the short rangeness of \(c\)-axis correlations,
they cannot be LR either. They then should be QLR, consistent with that demonstrated numerically.  

The peculiar transport properties observed in Bi\(_2\)Sr\(_2\)CaCu\(_2\)O\(_{8+y}\) 
when magnetic fields and currents are both parallel to the CuO\(_2\) layers are able to
be explained in terms of the \(B-T\) phase diagram, since dissipation at small currents should be 
governed by the equilibrium properties.  In the KT phase, isotropic helicity moduli with the
universal jump at the transition point are responsible to the isotropic, and therefore  
{\it Lorentz-force-independent}, dissipation and the power-law non-Ohmic I-V characteristics.  
In the 3D lattice phase,  which is realized at low magnetic fields for
Bi-based materials or up to quite strong fields in YBCO materials, larger dissipations should be observed at 
transverse currents than parallel ones.  

The \(B-T\) phase diagram in Fig.\ref{fig:pd} is consistent with several recent experiments.
By detecting the oscillation in the \(c\)-axis flux flow of Josephson vortices in 
Bi\(_2\)Sr\(_2\)CaCu\(_2\)O\(_{8+y}\), Ooi and Hirata succeeded in measuring
the phase boundary on which the 3D triangular lattice softens \cite{Ooi}.  The resultant curve has the same 
shape formed by \(T_{\times}(B)\) at high fields and \(T_m(B)\) at low fields, with a kink corresponding to the
multicritical point.  In the same material, Mirkovic {\it et al.} observed that the sharp drop of the resistivity 
associated with the first-order vortex lattice melting at low magnetic fields is suppressed into a smooth one 
when the magnetic field is elevated to about 1T, suggesting a continuous phase transition
\cite{Mirkovic}.  The steep normal to superconductivity phase boundary at high magnetic fields observed by 
Lundqvist {\it et al.} \cite{Lundqvist} is in accordance with the lower bound \(T^{\rm bare}_{\rm KT}\) on 
\(T_{\rm KT}(B)\), although the KT features are still not available experimentally.

Recently Kakeya {\it et al.} observed two plasma modes when applying a magnetic field parallel to the 
{\it ab} plane, of frequency higher (lower) than that of zero magnetic field and 
increasing (decreasing) with the magnetic field \cite{Kakeya}.   While the high branch is caused by the
resonance between the electromagnetic field and coherent motion of interlayer Josephson vortices, the low 
branch is asigned to the shear vibration of Josephson vortex lattice \cite{Koyama}.   Since the KT phase 
proposed in the present work is characterized by vanishing interlayer shear modulus, the low plasma mode
should disappear and the high mode is still observable as the magnetic field or temperature approach the phase
boundary \(T_{\times}\) from the 3D lattice phase.  Therefore, the Josephson plasma phenomenon provides
a powerful technique to test our phase diagram.

Since the multicritical field is approximately 50T for YBa\(_2\)Cu\(_3\)O\(_{7-\delta}\) presuming 
\(\gamma=8\) and \(d=12\AA\), the normal to superconductivity phase transition should be first order 
at magnetic fields available in laboratory to date.  This is consistent with a recent measurement on the 
specific heat by Schilling {\it et al.} up to 10T \cite{Schilling}.  It is revealed
that the phase boundary is smooth in accordance with the 3D anisotropic GL theory, and thus the system is 
essentially the same as the Abrikosov (or pancake) vortex system.  A meandering phase boundary was observed in 
transport measurements by Gordeev {\it et al.} in the same family of materials and field regime \cite{Zhukov}.
The difference might be quantitative, or may be the result of different experimental techniques.  It is 
noticed that our phase diagram is not to scale at low fields.

It is interesting to ask whether the phase diagram of interlayer Josephson vortices can be derived from a 
Lindemann type theory.  In a
Lindemann theory a lattice is supposed to melt when the thermal average of displacement
field exceeds a certain fraction of inter-vortex distance.  This picture has been useful in understanding the 
melting of pancake vortex lattice (or Bragg glass).  One should be very careful when
applying the Lindemann theory to the melting of interlayer Josephson vortex lattice, since
thermal fluctuations are very anisotropic especially at strong magnetic fields above
the multicritical point.  While thermal fluctuations in the \(c\) direction are pinned completely below
\(T_{\times}\), those in \(ab\) planes diverge as temperature approaches it from below.
Layers of Josephson vortices slide easily over each other in the intermediate phase.
A naive Lindemann theory is clearly not applicable.  

It is important to notice the differences among the present system and smectic liquid crystals
with or without an external field.  Ideal smecitc-A liquid crystals have liquid-like correlations in
two dimensions and a solid like periodic modulation of the density along the third direction.  The
elastic free-energy functional for a smectic possesses the so-called Landau-Peierls instability
\cite{Landau2,Peierls,deGennes}, {\it i.e.} suppression of in-plane quadratic first derivatives.
When a magnetic or dielectric field is applied, the layer normal is energetically confined
in a predetermined plane.   The quadratic first derivative is then suppressed in only one direction.
Systems in this group were called {\it planar layered}, and the possible phases and phase diagrams were 
addressed by Grinstein {\it et al.}  \cite{Grinstein}.  The interlayer
Josephson-vortex system in magnetic fields parallel to the layers is {\it polar layered}, in which the
normal of Josephson-vortex layers is along the crystalgraphic \(c\) axis.   There is no
Landau-Peierls like instability left any more in the present system.   The difference in symmetry results in
the different phases, as clearly indicated by the simulated structure factors.  It is interesting
to observe the similarity between the last two layered systems, namely both of them contain a multicritical 
point in their phase diagrams. 

\begin{acknowledgments}
The author would like to thank L. Balents, G. Baskaran, G. Blatter, L. Bulaevskii, Q.-H. Chen, J. Clem, G. Crabtree, 
K. Hirata, T. Ishida, K. Kadowaki, A. Koshelev, W. Kwok, S. Miyashita, D. Nelson, Y. Nonomura, A. Tanaka, 
V. Vinokur, U. Welp, and A. Zhukov for stimulating discussions.  
Simulations are performed on the Numerical Materials Simulator (SX-5) of National Institute for
Materials Science, Japan.  This work was partially supported by Japan Society for the
Promotion of Science (Grant-in-Aid for Scientific Research (C) No. 15540355).

\end{acknowledgments}

\end{document}